\documentclass[11pt]{article}
\usepackage[utf8]{inputenc}
\usepackage{amsmath}
\usepackage{amsfonts}
\usepackage{amssymb}
\usepackage{graphicx}
\usepackage{siunitx}
\usepackage{xcolor}
\usepackage{cite}
\usepackage{tikz}
\usepackage{bm,dsfont} % bbm package uses Type 3 fonts - avoid!
\usepackage{float}
\usepackage{pgfplots}
\usepgfplotslibrary{fillbetween}
\usepackage{url}
\usepackage{siunitx}
\usepackage{hyperref}
\usepackage{xcolor}
\usepackage{array,booktabs}
\usepackage{tikz-cd}
\hypersetup{linktocpage,pdfauthor={Omar},colorlinks=true,urlcolor=purple,linkcolor=purple,citecolor=purple}

\newcommand{\lambdabar}{{\mkern0.75mu\mathchar '26\mkern -9.75mu\lambda}}
\usetikzlibrary{
	babel,
	quotes,
	arrows.meta,
	shapes.symbols,
	shadows.blur,
}

\tikzset{
	> = LaTeX,
	pics/lisa/.style = {
		code = {
			\draw [densely dashed, thick] (0, 1) -- (210:1) -- (330:1) -- cycle;
			\foreach \th in {90, 210, 330} {
				\draw [thick, fill = white] (\th:1) circle [radius = 0.2];
			}
		}
	}
}

\definecolor{lime}{HTML}{A6CE39}
\DeclareRobustCommand{\orcidicon}{
	\begin{tikzpicture}
	\draw[lime, fill=lime] (0,0) 
	circle [radius=0.16] 
	node[white] {{\fontfamily{qag}\selectfont \tiny ID}};
	\draw[white, fill=white] (-0.0625,0.095) 
	circle [radius=0.007];
	\end{tikzpicture}
	\hspace{-2mm}
}

\foreach \x in {A, ..., Z}{%
	\expandafter\xdef\csname orcid\x\endcsname{\noexpand\href{https://orcid.org/\csname orcidauthor\x\endcsname}{\noexpand\orcidicon}}
}

\numberwithin{equation}{section}

\begin{document}
	
\title{\huge \textbf{Gravitational Faraday Effect induced by Dark Matter Spin}}
\author{
Francisco Barriga$^{1}$\thanks{fbarriga2017@udec.cl},
Fernando Izaurieta$^{2}$\thanks{fernando.izaurieta@uss.cl}\orcidF{},
Samuel Lepe$^{3}$\thanks{samuel.lepe@pucv.cl},
Paola Meza$^{2}$\thanks{paola.meza@uss.cl}, \\
Jethzael Mu\~{n}oz$^{1}$\thanks{jemunoz2017@udec.cl}, 
Cristian Quinzacara$^{2}$\thanks{cristian.quinzacara@uss.cl}\orcidQ{},
Omar Valdivia$^{4,5}$\thanks{ovaldivi@unap.cl}\orcidA{}.
\bigskip \\
%EndAName
{\small \textit{$^{1}$ Departamento de F\'{\i}sica, Universidad de Concepci\'on, casilla 160-C, Concepci\'on, Chile.}}\\
{\small \textit{$^{2}$ Facultad de Ingener\'{\i}a, Arquitectura y Dise\~no, Universidad San Sebasti\'an, Concepci\'on, Chile}}\\
{\small \textit{$^{3}$ Instituto de F\'{\i}sica, Facultad de Ciencias, Pontificia Universidad Cat\'olica de Valpara\'{\i}so, Avenida Brasil 2950, Valpara\'{\i}so, Chile.}}\\
{\small \textit{$^{4}$ Instituto de Ciencias Exactas y Naturales ICEN, Universidad Arturo Prat, Iquique, Chile.}}\\
{\small \textit{$^{5}$ Facultad de Ciencias FdeC, Universidad Arturo Prat, Iquique, Chile.%
}}
}
\maketitle
%%%In this paper, we explore an alternative approach to probing the fundamental properties of dark matter through gravitational interactions, focusing on its spin tensor. We investigate the impact of dark matter spin on the propagation of gravitational waves (GWs) within the context of Einstein-Cartan-Sciama-Kibble (ECSK) gravity, a theory that extends General Relativity by incorporating spacetime torsion. Specifically, w
\abstract{
\noindent We show that the spin of dark matter induces a gravitational analog of the electromagnetic Faraday effect, where the polarization of gravitational waves undergoes a rotation as they propagate through a dark matter halo with a non-vanishing axial spin tensor (hypermomentum). An expression for the gravitational rotation angle is provided, which is analogous to the Faraday rotation in optics, and evaluate its significance in astrophysical settings. Although the effect is expected to be small under current observational constraints, we discuss its potential importance in the early universe.}% and its relevance to the Hubble tension. Our results provide a novel mechanism for probing dark matter's spin properties through gravitational wave observations, opening a new avenue for exploring the dark sector using multimessenger astrophysics.}
%Samuel Lepe^{3‡}Instituto de Física, Facultad de Ciencias, Pontificia Universidad Católica de Valparaíso, Avenida Brasil 2950, Valparaíso, Chile, samuel.lepe@pucv.cl

%\end{abstract}
\newpage
%https://www.overleaf.com/project/620eb4d24c554ebf6e43f914
\tableofcontents
\newpage

%\begin{document}
\maketitle
\flushbottom
\section{Introduction}\label{Sec_Intro}

The nature of dark matter continues to present an ever-deepening mystery. Despite the incredible sensitivity of numerous clever experiments~\cite{XENON:2023cxc,XENON:2024ijk,LZ:2022lsv,LZ:2024psa,Pandey:2024dcd} designed to detect various dark matter candidates—such as WIMPs, axions, and other extensions to the Standard Model—none have yet yielded any conclusive results. If dark matter is indeed a particle, we must confront what has long been dubbed the \textquotedblleft Nightmare Scenario" in particle physics~\cite{Cho:2007cb,Bertone:2011kb}: dark matter may interact solely through gravity, making their detection in any foreseeable particle physics experiment impossible.

For this reason, we lack information on the basic features of dark matter; for instance, we do not have the slightest evidence pointing to whether dark matter is fermionic or bosonic. 

Given these challenges, it becomes imperative to explore whether gravity itself can serve as a probing tool for dark matter features~\cite{Tambalo:2022wlm,Ezquiaga:2020dao}. In this paper, we demonstrate how gravitational wave (GW) propagation is subtly influenced by interactions with dark matter~\cite{Siddhartha:2019yjm}, specifically through its spin tensor or hypermomentum~\cite{Hehl:1976kt,Hehl:1976kv}. In section~\ref{Sec_Faraday_Grav}, we reveal that an axial dark matter spin tensor impacts GWs in a manner analogous to the well-established Faraday rotation effect, resulting in a polarization rotation angle
\begin{equation}
	\Theta_{\mathrm{grav}} \left( \eta \right) = \kappa_{4}\int_{\eta=0}^{\eta} S^{0} \, \mathrm{d}\ell,
\end{equation}
where dark matter acts as a transparent dielectric, $\kappa_{4}=8\pi G$ plays the role of the Verdet constant, and the temporal component of the axial spin tensor $S^{0}$ mimics the parallel magnetic field component $B_{\parallel}$ in the electromagnetic analogy.

In section~\ref{Sec_Conclusions}, we evaluate the significance of this effect. Although it may be weak, it provides an avenue to gain insights into the fundamental nature of dark matter: A nonvanishing spin tensor would indicate a representation of the Lorentz group, suggesting that dark matter behaves as a hyperfluid and potentially possesses a fermionic nature.

The spin tensor naturally emerges when we vary the matter Lagrangian with respect to the affine degrees of freedom, much like the stress-energy tensor arises from variations with respect to the metric. Therefore, when considering non-vanishing spin tensors as a gravitational source, particularly in the context of fermionic dark matter, the most straightforward theoretical framework is Einstein-Cartan-Sciama-Kibble (ECSK) gravity, as detailed in section~\ref{Sec_ECSK}. ECSK extends General Relativity (GR) by treating metricity and affinity as independent degrees of freedom, thus relaxing the torsionless condition. In this framework, spin generates torsion in a manner analogous to how energy induces curvature. Of course, there are many other possibilities~\cite{Blagojevic:2013xpa,Hehl1980,Hehl76,BeltranJimenez2019,Addazi:2024rzo}, but ECSK is the most simple and closer to GR.

Numerous approaches exist for studying gravitational waves (GWs) within the context of Riemann-Cartan (RC) geometry (see refs.~\cite{Obukhov:2017pxa,Jimenez-Cano:2022arz,Jimenez-Cano:2020chm,Jimenez-Cano:2020lea,Ranjbar:2024bip,deAndrade:2021qew,Bluhm:2023kph,Ezquiaga:2017ner,Blagojevic:2017ssv,Bahamonde:2021dqn}). In section~\ref{Sec_GW_in_ECSK}, we provide a concise review of an approach developed in refs.~\cite{Izaurieta:2019dix,Barrientos2019,Barrientos:2019msu,Elizalde2023}, which allows for the analysis of wave propagation in RC geometries through an eikonal expansion, closely resembling the standard procedure in GR. The analysis to subleading order in the eikonal expansion leads directly to the gravitational analog of Faraday rotation, as discussed in sections~\ref{Sec_Faraday_Grav} and~\ref{Sec_Conclusions}.

The direct measurement of torsion is an old problem, and it still remains elusive~\cite{Hehl:1971vek}. It is notoriously difficult to detect for several reasons: it does not interact with classical particles~\cite{Hehl:2013qga,Puetzfeld:2014sja}, and, from an observational pseudo-Riemannian perspective, it could easily be mistaken for an extra dark species. This challenge has inspired numerous models where torsion plays a key role as a component of dark matter and dark energy~\cite{Barker:2020gcp,Alexander:2019wne,Alexander:2020umk,Tilquin:2011bu,Poplawski:2011jz,Poplawski:2011qr,Poplawski:2012qy,Unger:2018oqo,Desai:2016,Magueijo:2019vmk,Izaurieta:2020xpk,Poplawski:2020pie,Kranas:2018jdc,Pereira:2019yhu,Pereira:2022cmu,Guimaraes:2020drj,Ivanov:2016xjm,Razina:2010bj,Palle:2014goa,Rani:2022jam,Bakry:2021ajt,Barker:2020elg,Kasem:2020wsp,Cabral:2020mzw,Huang:2024ujj,Vicente:2024puv,Pervaiz:2023rkk,Usman:2023nww,Liu:2023znv,Benisty:2022lhr,Benisty:2021sul,Chervon:2023gio,deAndrade:2024yby,Gao:2024gat,Capozziello:2024lsz}, addressing a range of issues from inflation to the Hubble tension. 
The core challenge, however, is our inability to falsify these models, as current observational constraints on torsion remain extremely loose~\cite{Kranas:2018jdc,Pereira:2019yhu,Pereira:2022cmu,Guimaraes:2020drj}.

In particle physics, the situation for detecting torsion appears equally discouraging. If matter couples minimally to gravity, torsion would only interact with fermions within the Standard Model, and even then, the interaction would be so weak that it lies far beyond the reach of current experiments~\footnote{see the discussion at the end of chapter 8 in ref.~\cite{SupergravityVanProeyen}}. Nonminimal coupling offers a more promising outlook~\cite{Puetzfeld:2014sja,Carroll:1994dq}, but as of yet, no evidence suggests that Standard Model particles couple non-minimally.

In physics, while aesthetic considerations often guide our choice of hypotheses, the ultimate aim is to make predictions that can, at least in principle, be tested through observation. For this reason, the search for observational signatures of torsion remains crucial. We must determine whether torsion exists, rather than dismissing it as a mere artifact of our mathematical formulations. Although the specific effect discussed in this article, as shown in section~\ref{Sec_Conclusions}, is subtle and challenging to measure, many other promising proposals could one day offer stronger constraints on torsion. The most likely places to find evidence are in high-density fermion or dark matter environments, such as the early universe or neutron stars~\cite{Jockel:2024fps}.
%%%%%
%%%%
%%

%\textcolor{red}{
%Recordar:\\
%Añadir referencias de quienes nos han citado/pedido citas en el pasado.\\
%AÑADAN SUS NÚMEROS DE PROYECTOS FONDECYT EN LOS AGRADECIMIENTOS.
%}

\section{Preliminaries and Notation}\label{Sec_Notation}

In standard GR, the metric tensor encapsulates both the metric and affine properties when we impose the torsionless condition. However, in RC geometry, without enforcing vanishing torsion, metricity and affinity become independent properties of a manifold. Within this framework, two approaches to mathematical language and notation are available. The first utilizes differential forms on an orthonormal basis, where the vielbein $1$-form $e^{a} = e^{a}{}_{\mu} \mathrm{d}x^{\mu}$ governs metricity, and the spin connection $1$-form $\omega^{ab} = \omega^{ab}{}_{\mu} \mathrm{d}x^{\mu}$ defines affinity. The second approach relies on the standard tensorial language within a coordinate basis, where the metric $g_{\mu\nu}$ and the affine connection $\Gamma^{\lambda}{}_{\mu\nu}$ describe the geometry. The choice between these notations is both physically and mathematically inconsequential, but some ideas and calculations are more convenient in a certain language. For this reason, we will adopt whichever is more suitable in subsequent sections to explain different ideas.

We consider a $4$-dimensional spacetime manifold $M$ with a signature $\eta = (-,+,+,+)$. We use lowercase Greek indices to represent elements of the coordinate basis of vectors $\{\partial_{\mu}\}$ and $1$-forms $\{\mathrm{d}x^{\mu}\}$. Similarly, lowercase Latin indices are employed for the orthonormal basis of vectors $\{\hat{e}_{a} = e^{\mu}{}_{a} \partial_{\mu}\}$ and $1$-forms $\{e^{a} = e^{a}{}_{\mu} \mathrm{d}x^{\mu}\}$. We denote the space of p-forms on $M$ as $\Omega^{p}\left(M\right)$.

The Lorentz curvature $R^{ab}$ and torsion $T^{a}$ $2$-forms are given by
\begin{align}
    &R^{ab} = \mathrm{d}\omega^{ab} + \omega^{a}{}_{c} \wedge \omega^{cb}, \\
    &T^{a} = \mathrm{D} e^{a} = \mathrm{d}e^{a} + \omega^{a}{}_{b} \wedge e^{b},
\end{align}
where $\mathrm{d} : \Omega^{p}(M) \rightarrow \Omega^{p+1}(M)$ represents the exterior derivative, and $\mathrm{D} = \mathrm{d} + \omega : \Omega^{p}(M) \rightarrow \Omega^{p+1}(M)$ denotes the Lorentz-covariant derivative.

We denote the torsionless version of an entity with a circle above it. For example, we express the spin connection as the sum of a torsion-free component and a contortion $1$-form
\begin{align}
    \omega^{ab} = \mathring{\omega}^{ab} + \kappa^{ab},\label{Eq_Omega_Contorsion}
\end{align}
which satisfy the following conditions:
\begin{align}
    &\mathrm{d}e^{a} + \mathring{\omega}^{a}{}_{b} \wedge e^{b} = 0\,,\\
    &T^{a} = \kappa^{a}{}_{b} \wedge e^{b}.
\end{align}

In ref.~\cite{Barrientos2017,Izaurieta:2019dix,Barrientos2022}, we developed the mathematical framework for efficiently handling wave propagation in RC geometries for any theory. Moreover, in refs.~\cite{Barrientos2019} and~\cite{Elizalde2023}, we expanded this formalism to analyze gravitational waves within these geometries.

In this article, we will employ a couple of key definitions from those previous works. First, we introduce the generalized contraction operator $\mathrm{I}^{a} : \Omega^{p}\left(M\right) \rightarrow \Omega^{p-1}\left(M\right)$, defined as\footnote{Here $*$ stands for the Hodge dual operator which maps $\Omega^{p}(M)$ into $\Omega^{d-p}(M)$-forms.}
\begin{equation}
    \mathrm{I}^{a} = -\ast\left(e^{a} \wedge \ast\right. ,
\end{equation}
which allows us to define a new Lorentz derivative $\mathcal{D}_{a} : \Omega^{p}\left(M\right) \rightarrow \Omega^{p}\left(M\right)$ as
\begin{equation}
    \mathcal{D}_{a} = \mathrm{I}_{a}\mathrm{D} + \mathrm{D}\mathrm{I}_{a}.
\end{equation}

Using the canonical relation $\partial_{\mu}e^{a}{}_{\nu} + \omega^{a}{}_{b\mu}e^{b}{}_{\nu} - \Gamma^{\lambda}_{\mu\nu}e^{a}{}_{\lambda} = 0$, we can demonstrate that $\mathcal{D}_{a}$ is closely linked to the standard pseudo-Riemannian derivative $\nabla_{\mu} = \partial_{\mu} + \Gamma_{\mu}$ through
\begin{equation}
    \mathcal{D}_{a} = e_{a}{}^{\mu}\nabla_{\mu} + \mathrm{I}_{a}T^{b} \wedge \mathrm{I}_{b}.
\end{equation}
Therefore, $\mathcal{D}_{a}$ and $\nabla_{\mu}$ coincide when acting on $0$-forms and in the torsionless case.

The operator $\mathcal{D}_{a}$ is relevant because $-\mathcal{D}^{a}\mathcal{D}_{a}:\Omega^{p}\left(M\right) \rightarrow \Omega^{p}\left(M\right)$ corresponds to the generalization for a RC geometry of the Beltrami/Bochner wave operator~\cite{Barrientos2022}.

\section{The Spin Tensor and Gravitational Waves}

\subsection{ECSK gravity}\label{Sec_ECSK}

There are a huge number of modified theories of gravity with non-vanishing torsion, and in general, in all of them (even when torsion does not propagate in a vacuum), torsion will alter the propagation of GWs amplitude and/or polarization. It is possible to use the framework we have created in Refs.~\cite{Izaurieta:2019dix,Barrientos2022} to study this effect through the eikonal approximation that we already use in standard GR~\cite{Maggiore:1900zz}.

To show how it works, let us focus in the simplest case: ECSK gravity with axial torsion. The 4-form Lagrangian
\begin{equation}
    \mathcal{L}^{\left(4\right)}=\mathcal{L}^{\left(4\right)}_{\mathrm{G}}+\mathcal{L}^{\left(4\right)}_{\mathrm{M}},
\end{equation}
consists of a gravitational piece, considering independent metric and affine degrees of freedom, (the canonical Einstein-Hilbert term with cosmological constant),
\begin{align}
    \mathcal{L}^{\left(4\right)}_{\mathrm{G}}\left(e,\omega,\partial\omega\right)=\frac{1}{\kappa_{4}}\left(\frac{1}{4}\epsilon_{abcd}R^{ab}\wedge e^{c}\wedge e^{d}-\frac{\Lambda}{4!} \epsilon_{abcd} e^{a}\wedge e^{b}\wedge e^{c}\wedge e^{d}  \right)% , + \mathcal{L}_{M}(e, \omega , \psi),
\end{align}
and a minimally coupled matter Lagrangian,
\begin{equation}
    \mathcal{L}^{\left(4\right)}_{\mathrm{M}}=\mathcal{L}^{\left(4\right)}_{\mathrm{M}}\left(e, \omega, \psi,\partial\psi\right).
\end{equation}

The independent variations of the matter Lagrangian with respect the vierbein and the spin connection define the stress-energy 1-form $\tau_{b}=\tau_{ab}e^{a}$ and the spin tensor 1-form $\sigma_{ab}=\sigma_{cab}e^{c}$,

\begin{align}
    \delta_{e}\mathcal{L}^{\left(4\right)}_{\mathrm{M}}&=-\ast\tau_{d}\wedge\delta e^{d},\\
    \delta_{\omega}\mathcal{L}^{\left(4\right)}_{\mathrm{M}}&=-\frac{1}{2}\delta\omega^{ab}\wedge\ast\sigma_{ab}.
\end{align}
leading us to the well-known ECSK field equations
\begin{align}
    \frac{1}{2}\epsilon_{abcd}R^{ab}\wedge e^{c}-\frac{\Lambda}{3!}\epsilon_{abcd}e^{a}\wedge e^{b}\wedge e^{c}&=\kappa_{4}*\tau_{d},\label{Eq_ECSK_e}\\
    \epsilon_{abcd}T^{c}\wedge e^{d}&= \kappa_{4}*\sigma_{ab}.\label{Eq_ECSK_w}
\end{align}

In a region where the spin tensor vanishes (e.g., as in vacuum,  for classical matter, and for Yang-Mills bosons), torsion does not propagate and the ECSK dynamics reduce to standard GR. In the context of Standard Model particles, torsion becomes relevant only in environments with extremely high fermion densities, such as those found in the very early universe. Consequently, for most practical applications, ECSK theory and GR are considered equivalent, as torsion effects are typically negligible under standard astrophysical and cosmological conditions. However, we should not forget the possibility that dark matter possesses a non-vanishing spin tensor, which could serve as a source of torsion. In such scenario, torsion would contribute as an additional dark source for the standard torsionless pseudo-Riemannian piece of geometry. Consequently, the phenomena we attribute to dark matter might actually result from the interplay of both torsion and conventional dark matter effects.

In four dimensions, the spin tensor can be decomposed into three irreducible components under the Lorentz group: vector, axial, and tensorial
\begin{equation}
\sigma_{\lambda \mu \nu}=\frac{2}{3}\left(  \Sigma_{\lambda \mu \nu}-\Sigma_{\lambda \nu \mu}\right)  +\frac{1}{3}\left(  g_{\lambda \mu}\Sigma_{\nu}-g_{\lambda \nu}\Sigma_{\mu}\right)  +2\sqrt{\left \vert g\right \vert }\epsilon_{\lambda \mu \nu \rho}S^{\rho}.
\end{equation}
The vector component is given by
\begin{equation}
\Sigma_{\nu}=\sigma^{\lambda}{}_{\lambda \nu},
\end{equation}
while the axial and tensorial components are
\begin{equation}
S^{\mu}=\frac{1}{2}\frac{1}{3!\sqrt{\left \vert g\right \vert }}\epsilon^{\mu \nu \rho \sigma}\sigma_{\nu \rho \sigma},
\end{equation}
\begin{equation}
\Sigma_{\lambda \mu \nu}=\frac{1}{2}\left(  \sigma_{\lambda \mu \nu}+\sigma_{\mu \lambda \nu}\right)  +\frac{1}{3!}\left(  g_{\nu \lambda}\Sigma_{\mu}+g_{\nu \mu}\Sigma_{\lambda}\right)  -\frac{1}{3}g_{\lambda \mu}\Sigma_{\nu},
\end{equation}
respectively.

While individual Standard Model fermions generate only an axial component (the chiral current), a dense, strongly interacting fermion plasma is effectively described by a Weyssenhoff fluid, which includes a tensorial component. Thus, when hypothesizing a non-vanishing spin tensor for dark matter—such as in the case of a non-interacting fermion particle—considering an axial spin tensor (and its corresponding torsion) becomes natural. This is because, for non-interacting particles, it remains unclear whether gravity alone can induce decoherence, as suggested by some studies on neutrinos (see ref.~\cite{IceCubeColb2024}). If decoherence does occur, it likely operates at a much slower rate than in Standard Model particles, opening the intriguing possibility of quantum effects manifesting on astrophysical scales, as seen in models like fuzzy dark matter or superfluid dark matter~\cite{Lee2018,Hui2017,Berezhiani2015,Khoury2016,Hossenfelder2017,Hossenfelder2019,Mistele2022}. In our case, we are considering the possibility of dark matter being a quantum hyperfluid, so it is reasonable to consider it having a purely axial, macroscopic dark matter spin tensor
\begin{equation}
\sigma_{\lambda \mu \nu}=2\sqrt{\left \vert g\right \vert }\epsilon_{\lambda \mu \nu \rho}S^{\rho}.
\end{equation}

An axial dark matter spin tensor offers significant advantages: it provides a parametrically simple explanation for the Hubble tension~\cite{Vagnozzi:2023nrq,Vagnozzi:2019ezj} while preserving the successes of $\Lambda$CDM (see refs.~\cite{Izaurieta:2020xpk,Koivisto:2023epd}). More broadly, ECSK torsion introduces several fascinating effects such as inflation-like behaviors (see refs.~\cite{Poplawski:2010kb,Desai:2016}), potential roles as dark matter or dark energy candidates, and a reduction of the cosmological constant problem to a discrepancy of just eight orders of magnitude when accounting for torsional contributions to quantum vacuum fluctuations (see ref.~\cite{Poplawski:2011qr}).

Given these compelling possibilities, it becomes crucial to identify a probe capable of distinguishing torsion, particularly axial torsion, from other dark sources. As we will show in the following sections, perhaps gravitational waves could serve this purpose. Specifically, axial torsion produces a gravitational analog of Faraday rotation, which we will explore in detail.

\subsection{Gravitational Waves in ECSK Gravity}\label{Sec_GW_in_ECSK}

In a Riemann-Cartan (RC) geometry, the spin connection $\omega^{ab}$ and the vielbein $e^{a}$ generally represent independent degrees of freedom. To analyze GWs within this framework, we introduce independent perturbation 1-forms:

\begin{align}
    &e^{a} \rightarrow \Bar{e}^{a}=e^{a}+\frac{1}{2}H^{a},\\
    &\omega^{ab} \rightarrow \Bar{\omega}^{ab}=\omega^{ab}+u^{ab}.
\end{align}

Analogous to the decomposition of the spin connection eq.~(\ref{Eq_Omega_Contorsion}), we find it advantageous to decompose the perturbation $u^{ab}$ in a similar fashion
\begin{equation}
    u^{ab}=U^{ab}\left(H\right) + V^{ab},
\end{equation}
where
\begin{equation}
    U^{ab}\left(H\right)=-\frac{1}{2}\left(\mathrm{I}_{a}\mathrm{D}H_{b}-\mathrm{I}_{b}\mathrm{D}H_{a}\right)+\mathcal{O}\left(H^{2}\right)
\end{equation}
captures the dependency on $H^{a}=H_{ab}e^{b}$ (see ref.~\cite{Izaurieta:2019dix} for a detailed treatment on RC perturbations with this notation). The remaining term, $V^{ab}$, represents an independent \textquotedblleft roton"~\cite{Boos:2016cey} or contortional perturbation for theories where torsion propagates in a vacuum. In ECSK theory, where torsion does not propagate in a vacuum, $V^{ab}$ can be resolved in terms of $H^{a}$ and the background torsion, in such a way that $V^{ab}=0$ when the background torsion vanishes.

To be considered a GW, a perturbation must satisfy the eikonal condition, as described in ref.~\cite{Maggiore:1900zz}. Specifically, the perturbations take the form
\begin{align}
    H^{a}&=\mathrm{e}^{i\theta}\mathcal{H}^{a},\label{Eq_Eikonal_H}\\
    V^{ab}&=\mathrm{e}^{i\theta}\mathcal{V}^{ab},
\end{align}
where $\theta$ is a rapidly varying phase over a small characteristic length $\lambda$, and its derivative defines the \textquotedblleft wave 1-form"

\begin{equation}
    k_{\mu}=\partial_{\mu}\theta\sim\frac{1}{\lambdabar}.
\end{equation}
The amplitudes $\mathcal{H}^{a}$ and $\mathcal{V}^{ab}$ vary slowly over a long characteristic length $L$. The eikonal condition is then expressed as

\begin{equation}
    \epsilon=\frac{\lambdabar}{L}\ll 1.
\end{equation}

Considering terms up to leading ($1/\epsilon^2$) and subleading ($1/\epsilon$) orders in the eikonal approximation, the perturbations of the ECSK field equations yield (see ref.~\cite{Elizalde2023})

\begin{equation}
    -\frac{1}{2}\left(\mathrm{I}_{n}\mathcal{D}_{a}\mathcal{D}^{a}H_{m}-\mathrm{I}_{n}[\mathcal{D}_{a},\mathcal{D}_{m}]H^{a}\right)+\left(\mathrm{I}_{n}\mathcal{D}_{a}-\mathcal{D}_{n}\mathrm{I}_{a}\right)V^{a}{}_{m}=0.
\end{equation}
At the leading order ($1/\epsilon^2$), this equation dictates the dispersion relation $k^{\lambda}k_{\lambda}=0$, indicating that GWs propagate along null curves, just as in GR. Furthermore, it ensures that

\begin{equation}
    k^{\mu}\mathring{\nabla}_{\mu}k^{\lambda}=0\,,\label{Eq_GW_Geodesic}
\end{equation}
implying that GWs in ECSK gravity propagate along null torsionless geodesics (rather than along null auto-parallels), regardless of the presence of background torsion. This result is crucial from an observational standpoint. Electromagnetic waves, like any Yang-Mills interaction, remain unaffected by torsion and thus propagate along null torsionless geodesics, independent of any torsional background. We see here that both, GWs and electromagnetic waves, travel at the same speed and along the same types of curves, irrespective of torsion, showing that ECSK torsion does not contradict multimessenger observations~\cite{Barrientos2019,PhysRevLett.119.161101,GBM:2017lvd,Monitor:2017mdv,Huerta2019,Bettoni:2016mij,Ezquiaga:2018btd}.

At the subleading order in the eikonal approximation ($1/\epsilon$), we gain insights into the propagation of amplitude and polarization. 
To encapsulate this information, we express $\mathcal{H}^{a}$ from eq.~(\ref{Eq_Eikonal_H}) as

\begin{equation}
    \mathcal{H}^{a}=\varphi P^{a}{}_{b}e^{b},
\end{equation}
where the polarization components $P_{ab}$ are complex numbers, normalized such that

\begin{equation}
    \bar{P}^{ab}P_{ab}=1,\label{Eq_PP=1}
\end{equation}
with the bar denoting complex conjugation. The scalar amplitude $\varphi$ is a real number and, together with the wave vector $k_{\mu}$, one can define the current density for the \textquotedblleft number of gravitons"
\begin{equation}
    J_{\mu}=\varphi^{2}k_{\mu}.
\end{equation}
which is in fact a conserved quantity when torsion carries only an axial component. 
Focusing on such scenario where dark matter is the source of axial torsion (see ref.~\cite{Elizalde2023}), we have
\begin{equation}
T_{\lambda\mu\nu}=2\kappa_{4}\sqrt{\left|g\right|}\epsilon_{\lambda\mu\nu\sigma}S^{\sigma}.
\end{equation}
Therefore, at the subleading order in the eikonal approximation, we find
\begin{align}
    \mathring{\nabla}_{\lambda}J^{\lambda}&=0,\label{Eq_DJ=0}\\
    k^{\lambda}\mathring{\nabla}_{\lambda}P_{\mu\nu}&=-\kappa_{4}\sqrt{\left|g\right|}k^{\lambda}S^{\sigma}\left(\epsilon_{\nu\rho\lambda\sigma}P^{\rho}{}_{\mu}+\epsilon_{\mu\rho\lambda\sigma}P^{\rho}{}_{\nu}\right).\label{Eq_kDP=T}
\end{align}
Equation~(\ref{Eq_DJ=0}) indicates that, while propagating along the torsionless null geodesic, the amplitude of the GW decays in the same manner as in standard GR. In contrast, in standard GR, the right-hand side of eq.~(\ref{Eq_kDP=T}) would vanish, implying that polarization is parallel transported by the GW along the torsionless null geodesic. However, when torsion is present, the right-hand side of~(\ref{Eq_kDP=T}) does not vanish, revealing that polarization evolves anomalously during propagation.

\section{Gravitational Faraday Effect}\label{Sec_Faraday_Grav}

To analyze the propagation of polarization as described by eq.~(\ref{Eq_kDP=T}), it is useful to work within an orthonormal frame. In this frame, the equation becomes
\begin{equation}
    k^{c}\mathring{\mathcal{D}}_{c}P_{ab}=-\kappa_{4}k^{c}S^{m}\left(\epsilon_{b n c m}P^{n}{}_{a}+\epsilon_{a n c m}P^{n}{}_{b}\right).\label{Eq_kDP-Ortonormal}
\end{equation}

We align the third direction of the orthonormal frame with the direction of GW propagation, imposing $k^{1} = k^{2} = 0$, and satisfying the condition
\begin{equation}
    -\left(k^{0}\right)^2 + \left(k^{3}\right)^2 = 0.
\end{equation}
In principle, in a theory that incorporates torsion, we must consider all six possible polarization modes of a GW. Therefore, the polarization can be decomposed as

\begin{equation}
    P_{ab} = p_{\left(I\right)}\mathbb{P}^{\left(I\right)}_{ab},\label{Eq_P=p_I_P_I}
\end{equation}
where the $p_{\left(I\right)}$ are complex scalars and $I=\left\{+,\times, b, l, x, y\right\}$ labels the standard $+$ and $\times$ modes, the breathing and longitudinal trace modes, and the $x$ and $y$ vector modes (we adopted the Einstein summation convention on the index $\left(I\right)$ to simplify notation). There are many choices for a basis in the space of polarizations, but without losing generality, we can chose an orthonormal basis in the polarization space,
\begin{equation}
    \bar{\mathbb{P}}^{ab}_{\left(J\right)}\mathbb{P}_{ab}^{\left(I\right)}=\delta^{I}_{J}.\label{Eq_P_orthonormal}
\end{equation}
In this case, equations~(\ref{Eq_P_orthonormal}) and~(\ref{Eq_PP=1}) lead us to the relation
\begin{equation}
\bar{p}^{\left(I\right)}p_{\left(I\right)} = 1.
\end{equation}
In Appendix~\ref{App_Polarization}, we provide an explicit orthonormal (and real) basis for the polarization space, that correspond to GW modes propagating along the third direction of the orthonormal frame.

In the following sections, we demonstrate that in ECSK gravity with a dark matter background featuring a non-vanishing axial spin tensor, only the standard polarization modes $\left\{+,\times\right\}$ persist. However, to maintain generality, we will initially consider all possible polarization modes, illustrating the procedure for more comprehensive cases before showing how ECSK dynamics ultimately restricts the allowed modes to $\left\{+,\times\right\}$.

Let us consider a simple but useful model assuming a weak torsion GW emission scenario ref.~\cite{Elizalde2023}, where the emission process is essentially the same as in standard GR\footnote{The GW emission process in ECSK has been deeply studied in refs.~\cite{Battista:2023znv,Battista:2022sci,Battista:2022hmv,Battista:2021rlh}. Departures from GR should be relevant for an extremely dense fermion plasma background.}. This means we are assuming a torsion background
which is too weak to affect the emission process of the gravitational wave in
an observable manner. While torsion effects may accumulate along the geodesic during propagation, the GW emission itself remains unaffected by torsion, just as in GR. This assumption is well-founded. In general, the effects of dark matter on GW emission are negligible at leading and subleading order in the eikonal limit. Since we are modeling torsion as part of the dark matter sector, torsion’s influence is even weaker in this context.

We describe the background geometry as given in eq.~(\ref{Eq_Omega_Contorsion}) and introduce

\begin{equation}
    \mathring{P}_{ab} = \mathring{p}_{\left(I\right)}P^{\left(I\right)}_{ab},
\end{equation}
as the polarization that would propagate in standard GR over the torsionless background represented by $\mathring{\omega}^{ab}$. Therefore, at leading and subleading orders in the eikonal approximation

\begin{equation}
    \mathring{p}_{\left(b\right)} = \mathring{p}_{\left(l\right)} = \mathring{p}_{\left(x\right)} = \mathring{p}_{\left(y\right)} = 0.\label{Eq_p_torsionless_nulos}
\end{equation}

Let us now define the affine geodesic parameter $\eta$, such that $\eta = 0$ marks the point of GW emission. The ECSK propagation of polarization modes along the geodesic is described as a deviation from the standard GR case

\begin{equation}
    p_{\left(A\right)}\left(\eta\right) = S^{\left(B\right)}{}_{\left(A\right)}\left(\eta\right)\mathring{p}_{\left(B\right)}\left(\eta\right),\label{Eq_p=Sp}
\end{equation}
where the matrix $S^{\left(B\right)}{}_{\left(A\right)}\left(\eta\right)$ encodes how torsion \textquotedblleft scrambles" the polarization modes as the GW propagates along the null geodesic parameterized by $\eta$. Owing to eq.~(\ref{Eq_p_torsionless_nulos}), there are obvious degeneracies in this matrix; without loss of generality we set $S^{\left(b\right)}{}_{\left(A\right)} = S^{\left(l\right)}{}_{\left(A\right)} = S^{\left(x\right)}{}_{\left(A\right)} = S^{\left(y\right)}{}_{\left(A\right)} = 0$.

In the weak torsion scenario, at the moment of emission $\eta = 0$, the only nonzero components are

\begin{equation}
    S^{\left(+\right)}{}_{\left(+\right)}\left(0\right) = S^{\left(\times\right)}{}_{\left(\times\right)}\left(0\right) = 1.\label{Eq_S_ini}
\end{equation}
Substituting eq.~(\ref{Eq_P=p_I_P_I}) into eq.~(\ref{Eq_kDP-Ortonormal}), and using eq.~(\ref{Eq_p=Sp}) along with the orthonormality of the polarization basis in eq.~(\ref{Eq_P_orthonormal}), we derive the propagation relation for the matrices $S^{\left(J\right)}{}_{\left(I\right)}$ along the null geodesic

\begin{equation}
k^{c}\left(  \mathcal{\mathring{D}}_{c}\mathring{p}_{\left(  J\right)}S^{\left(  J\right)  }{}_{\left(  I\right)  }+\mathring{p}_{\left(  J\right)}\mathcal{\mathring{D}}_{c}S^{\left(  J\right)  }{}_{\left(  I\right)}+p_{\left(  J\right)}\left(  \mathring{\omega}_{abc}+\kappa_{4}S^{s}\epsilon_{abcs}\right)  \left[  \mathbb{\bar{P}}_{\left(  I\right)},\mathbb{P}^{\left(  J\right)  }\right]  ^{ab}\right)  = 0\,,\label{Eq_DS-1}
\end{equation}
where the torsionless spin connection components are given by $\mathring{\omega}_{ab} = \mathring{\omega}_{abc} e^{c}$.

The torsionless polarization $\mathring{P}_{ab} = \mathring{p}_{\left(I\right)}\mathbb{P}_{ab}^{\left(I\right)}$ satisfies the equation $k^{c}\mathcal{\mathring{D}}_{c}\mathring{P}_{ab} = 0$ along the geodesic. In terms of the polarization components, this leads to the following propagation relation for $\mathring{p}_{\left(I\right)}$

\begin{equation}
k^{c}\left(  \mathcal{\mathring{D}}_{c}\mathring{p}_{\left(I\right)} + \mathring{\omega}_{abc}\mathring{p}_{\left(J\right)}\left[\mathbb{\bar{P}}_{\left(I\right)},\mathbb{P}^{\left(J\right)}\right]^{ab}\right)  = 0.\label{Eq_Dp_torsionless}
\end{equation}
Substituting eq.~(\ref{Eq_Dp_torsionless}) into eq.~(\ref{Eq_DS-1}), and using the geodesic tangent vector $t^{\mu}=\frac{\mathrm{d}X^{\mu}}{\mathrm{d}\eta}\propto k^{\mu}$, we obtain the following equation for the propagation of $S^{\left(A\right)}{}_{\left(B\right)}$ along the geodesic

\begin{equation}
\frac{\mathrm{d}S^{\left(A\right)}{}_{\left(B\right)}}{\mathrm{d}\eta} + t^{c}\mathring{\omega}_{abc}\left(S^{\left(A\right)}{}_{\left(C\right)}\left[\mathbb{\bar{P}}_{\left(B\right)},\mathbb{P}^{\left(C\right)}\right]^{ab} - \left[\mathbb{\bar{P}}_{\left(C\right)},\mathbb{P}^{\left(A\right)}\right]^{ab}S^{\left(C\right)}{}_{\left(B\right)}\right) + S^{\left(A\right)}{}_{\left(C\right)}\kappa_{4}S^{s}t^{c}\epsilon_{abcs}\left[\mathbb{\bar{P}}_{\left(B\right)},\mathbb{P}^{\left(C\right)}\right]^{ab}=0.\label{Eq_DS_Gen}
\end{equation}

Let us study this equation for a simple but representative case, by considering a GW that goes through a null radial geodesic in a dark matter distribution with spherical symmetry. The spherically symmetric background created by the dark matter halo corresponds to a tetrad of the form ($c=1$)

\begin{align}
e^{0} & = \mathrm{e}^{\alpha \left( t,r\right)} \, \mathrm{d}t, \\
e^{1} & = r \, \mathrm{d}\theta, \\
e^{2} & = r \sin \theta \, \mathrm{d}\phi, \\
e^{3} & = \mathrm{e}^{\beta \left( t,r\right)} \, \mathrm{d}r,
\end{align}

A corresponding axial spin tensor where the only non-vanishing component is $S^{0} \neq 0$ is also spherically symmetric. This background geometry can describe a spherical dark matter halo but also a FLRW background for GW propagation. The non-vanishing components of the torsionless spin connection are:

\begin{align}
\mathring{\omega}_{03} & = -\left( \mathrm{e}^{-\beta}\alpha^{\prime} e^{0} + \mathrm{e}^{-\alpha} \dot{\beta} e^{3} \right),\label{Eq_Omega_Ini} \\
\mathring{\omega}_{31} & = -\frac{1}{r} \mathrm{e}^{-\beta} e^{1}, \\
\mathring{\omega}_{32} & = -\frac{1}{r} \mathrm{e}^{-\beta} e^{2}, \\
\mathring{\omega}_{12} & = -\frac{1}{r \tan \theta} e^{2}.\label{Eq_Omega_Fin}
\end{align}
By inspecting eq.~(\ref{Eq_DS_Gen}) with the components eqs.~(\ref{Eq_Omega_Ini}--\ref{Eq_Omega_Fin}) and considering that $t^{1}=t^{2}=0$, we observe that it simplifies to

\begin{equation}
\frac{\mathrm{d}S^{\left(A\right)}{}_{\left(B\right)}}{\mathrm{d}\eta} - 2S^{\left(A\right)}{}_{\left(C\right)}\kappa_{4} S^{0} t^{3} \left[ \mathbb{\bar{P}}_{\left(B\right)}, \mathbb{P}^{\left(C\right)} \right]^{12} = 0.\label{Eq_dS/dn}
\end{equation}

When considering the polarization basis commutators (Appendix~\ref{App_Polarization}), we find that the only polarization commutators that could contribute to this equation are $\left[ \mathbb{P}^{\left(+\right)}, \mathbb{P}^{\left(\times\right)} \right]$ and $\left[ \mathbb{P}^{\left(x\right)}, \mathbb{P}^{\left(y\right)} \right]$. However, including the initial conditions in eq.~(\ref{Eq_S_ini}), it is straightforward to prove that only the commutator $\left[ \mathbb{P}^{\left(+\right)}, \mathbb{P}^{\left(\times\right)} \right]$ survives. Solving the differential system in eq.~(\ref{Eq_dS/dn}), we find that the only non-vanishing components of $S^{\left(A\right)}{}_{\left(B\right)}$ correspond to the transverse rotation matrix

\begin{align}
S^{\left( + \right)}{}_{\left( + \right)} & = \cos 2\Theta_{\mathrm{grav}}, & S^{\left( + \right)}{}_{\left( \times \right)} & = -\sin 2\Theta_{\mathrm{grav}}, \\
S^{\left( \times \right)}{}_{\left( + \right)} & = \sin 2\Theta_{\mathrm{grav}}, & S^{\left( \times \right)}{}_{\left( \times \right)} & = \cos 2\Theta_{\mathrm{grav}},
\end{align}
where the rotation angle is given by

\begin{equation}
\Theta_{\mathrm{grav}} \left( \eta \right) = \kappa_{4}\int_{\eta=0}^{\eta} S^{0} \, \mathrm{d}\ell, \label{Eq_Theta_GW}
\end{equation}
with $\mathrm{d}\ell$ representing the spatial element along the GW path. The integral runs from the point of emission to the point of observation.

The net effect is a rotation of the polarization with respect to the GR case, where instead of the GR polarization $\mathring{P}_{ab} = \mathring{p}_{\left(+\right)} \mathbb{P}_{ab}^{\left(+\right)} + \mathring{p}_{\left(\times\right)} \mathbb{P}_{ab}^{\left(\times\right)}$ we get the rotation
\begin{equation}
P_{ab} = \mathring{p}_{\left(+\right)} \left( \cos 2\Theta_{\mathrm{grav}} \mathbb{P}_{ab}^{\left(+\right)} - \sin 2\Theta_{\mathrm{grav}} \mathbb{P}_{ab}^{\left(\times\right)} \right) + \mathring{p}_{\left(\times\right)} \left( \cos 2\Theta_{\mathrm{grav}} \mathbb{P}_{ab}^{\left(\times\right)} + \sin 2\Theta_{\mathrm{grav}} \mathbb{P}_{ab}^{\left(+\right)} \right), \label{Eq_Polarization_Rotation}
\end{equation}
with the sign of $S^{0}$ determining the chirality of the rotation. A rotation angle of $\Theta_{\mathrm{grav}} = \pi / 4$ swaps the $\left( + \right)$ and $\left( \times \right)$ polarization modes.

The effect described by eq.~(\ref{Eq_Polarization_Rotation}) bears a striking resemblance to the well-known Faraday rotation in optics (\cite{FaradaysDiary4,Prati_Faraday:2003}). In standard Faraday rotation, the polarization of an electromagnetic wave rotates when a magnetic field component $B_{\parallel}$ is parallel to the light path through a dielectric medium. The rotation angle $\theta_{\mathrm{Faraday}}$ is given by
\begin{equation}
\theta_{\mathrm{Faraday}}\left(\eta\right) = \mathcal{V} \int_{\eta=0}^{\eta} B_{\parallel} \, \mathrm{d}\ell, \label{Eq_Faraday}
\end{equation}
where $\mathcal{V}$ is the Verdet parameter of the medium, and the integral is evaluated along the light path. The similarity between eqs.~(\ref{Eq_Faraday}) and~(\ref{Eq_Theta_GW}) is remarkable. In some sense, the constant $\kappa_{4}=8\pi G$ plays the role of the dielectric Verdet parameter, and the spin tensor axial term $S^{0}$ plays a role analogous to the magnetic field $B_{\parallel}$; the dark matter spin tensor seems to act as the gravitational analog of a transparent dielectric medium. However, the physical mechanisms behind the two effects are fundamentally different. In the case of Faraday rotation, the dispersion relation (leading order in the eikonal limit) is modified by circular birefringence: left- and right-handed circularly polarized light propagates at slightly different speeds, producing the net rotation. By contrast, the gravitational rotation in eq.~(\ref{Eq_Polarization_Rotation}) arises from torsion-induced effects at subleading order in the eikonal limit, with no change in GW propagation speed. As a consequence, the Verdet parameter (and $\theta_{\mathrm{Faraday}}$) is highly dependent on the wavelength, while $\Theta_{\mathrm{grav}}$ is not.

%\subsection{Estimations of $\Theta_{\mathrm{grav}}$ for Astrophysical Scenarios}

\section{Conclusions: Testing and Future Developments.}\label{Sec_Conclusions}

When formulating a scientific hypothesis, the epistemological etiquette indicates we should try to provide a prediction that could be tested in some experiment. For the gravitational Faraday effect proposed here, both aspects (prediction and testing) prove challenging despite being a simple and general phenomenon in the context of ECSK theory.

For starters, in laboratory conditions the electromagnetic Faraday effect can be easily measured by comparing the input and output polarization states. However, at astrophysical scales, the initial polarization state of a GW at emission is unobservable for practical purposes, making a direct measurement of $\Theta_{\mathrm{grav}}$ highly unlikely in the near future. Nonetheless, indirect approaches, such as examining cases of GW lensing and interference across different trajectories, could provide a viable solution (see refs.~\cite{PhysRevD.100.064028,Harikumar:2023gzh} for interesting approaches for studying changes in polarization through GW interference). More work along these lines will be presented elsewhere. 

While equation (\ref{Eq_Theta_GW}) establishes a theoretical framework, predicting $\Theta_{\mathrm{grav}}$ within the ECSK theory remains a challenging task. Current observational constraints on the dark matter spin tensor (and torsion), and particularly in its potential role as an extra dark matter or dark energy contribution, are still extremely loose. As a result, estimates of $\Theta_{\mathrm{grav}}$ heavily rely on the specific model adopted for the dark sector; see refs.~\cite{Morawetz:2023jll,Li:2023ubd,Yun:2024xzk,Izaurieta:2020kuy,Cid:2017wtf,Cruz:2020hkh,Izaurieta:2020vyh} to have a view of different ways torsion can play a role as part of the dark sector during cosmic evolution.

To explore this further, let us briefly review propagation on a FLRW geometry. Let us consider a FLRW metric with a flat spatial section, an axial spin tensor along the temporal direction, and a matter Lagrangian including baryons and dark matter. Then, the field equations (\ref{Eq_ECSK_e}, \ref{Eq_ECSK_w}) predict the axial ECSK-Friedmann equations ($c=1$)
\begin{align}
3H^{2}-\kappa_{4}\left(  \rho_{\mathrm{b}}+\rho_{\mathrm{D}}\right)   & =0 ,\\
\frac{\mathrm{d}}{\mathrm{d}t}\rho_{\mathrm{b}}+3H\left(  \rho_{\mathrm{b}}+p_{\mathrm{b}}\right)    & =0,\\
\frac{\mathrm{d}}{\mathrm{d}t}\rho_{\mathrm{D}}+3H\left(  \rho_{\mathrm{D}}+p_{\mathrm{D}}\right)    & =0,
\end{align}
where \textquotedblleft b\textquotedblright\ refers to baryonic matter, and \textquotedblleft D\textquotedblright\ denotes the dark sector, given by
\begin{align}
\rho_{\mathrm{D}}  & =\rho_{\mathrm{DM}}+3\kappa_{4}\left[  S^{0}\right]^{2}+\frac{\Lambda}{\kappa_{4}},\label{Eq_rho_D}\\
p_{\mathrm{D}}  & =p_{\mathrm{DM}}-\kappa_{4}\left[  S^{0}\right]  ^{2}-\frac{\Lambda}{\kappa_{4}},\label{Eq_p_D}
\end{align}
where \textquotedblleft DM\textquotedblright\ stands for dark matter. Here, the axial spin term $\left[ S^{0}\right]^{2}$ acts as an additional dark component. To solve this system, we require an equation of state that links $S^{0}$ with $\rho_{\mathrm{DM}}$. However, due to our limited understanding of dark matter, this is precisely what we lack. Thus, several models compatible with current observations are possible (e.g., see refs.~\cite{Izaurieta:2020xpk,Yun:2024xzk,Koivisto:2023epd}). % no se te olvide sacar las XXX

What is clear, though, is that the \emph{observed} effects of dark matter and dark energy could stem from a combination of a \textquotedblleft bare\textquotedblright\ term, $\frac{\Lambda}{\kappa_{4}} + \rho_{\mathrm{DM}}$, amplified by $\left[ S^{0}\right]^{2}$ through equations (\ref{Eq_rho_D}, \ref{Eq_p_D}). It is feasible to develop models that slightly deviate from the $\mathrm{\Lambda CDM}$ paradigm, where $\left[ S^{0}\right] ^{2} \ll \rho_{\mathrm{DM}}$, and the small, additional negative pressure caused by $\left[ S^{0}\right]^{2}$ resolves the Hubble tension (see ref.~\cite{Izaurieta:2020xpk}). Furthermore, models can be constructed where $\left[ S^{0}\right]^{2}$ plays a pivotal role in explaining dark energy (see ref.~\cite{Yun:2024xzk}).

To estimate the order of magnitude of $\Theta_{\mathrm{grav}}$ and evaluate the significance of this effect, we turn to the model proposed in ref.~\cite{Izaurieta:2020xpk}. In this reference, based on equations (\ref{Eq_rho_D}, \ref{Eq_p_D}), we introduced an equation of state of the form
\begin{equation}
S^{0}=-\alpha_{\mathrm{Y}}\sqrt{\frac{1}{3\kappa_{4}}\rho_{\mathrm{DM}}},
\end{equation}
where $\alpha_{\mathrm{Y}}$ is the so-called \textquotedblleft spintropic" constant, and we consider standard cold dark matter $p_{\mathrm{DM}}=0$. In this model, the axial spin generates an effective dark matter density and pressure in the FLRW equations, given by
\begin{align}
\rho_{\mathrm{eff}}  & =\left(  1+\alpha_{\mathrm{Y}}^{2}\right)\rho_{\mathrm{DM}},\\
p_{\mathrm{eff}}  & =-\frac{1}{3}\alpha_{\mathrm{Y}}^{2}\rho_{\mathrm{DM}}=\omega_{\mathrm{eff}}\rho_{\mathrm{eff}},
\end{align}
where the effective negative barotropic constant is
\begin{equation}
\omega_{\mathrm{eff}}=-\frac{1}{3}\frac{\alpha_{\mathrm{Y}}^{2}}{1+\alpha_{\mathrm{Y}}^{2}}.
\end{equation}
This constant arises in the FLRW equations despite the cold dark matter having a bare barotropic parameter of $\omega_{\mathrm{DM}}=0$. Refs.~\cite{Poplawski:2019tub,Izaurieta:2020xpk} showed that a small effective barotropic parameter, $\omega_{\mathrm{eff}}\approx10^{-2}$ (corresponding to $\alpha_{\mathrm{Y}}=0.183$), is sufficient to resolve the Hubble parameter tension.

In this framework, the gravitational rotation angle is given by:
\begin{equation}
\Theta_{\mathrm{grav}}=-\mathrm{sgn}\left(  \alpha_{\mathrm{Y}}\right)\int_{\eta=0}^{\eta}\sqrt{-\kappa_{4}\omega_{\mathrm{eff}}\rho_{\mathrm{eff}}}\, \mathrm{d}\ell.
\end{equation}

When calculating the rate of change of the rotation angle
\begin{equation}
\frac{\mathrm{d}\Theta_{\mathrm{grav}}}{\mathrm{d}\ell}=-\mathrm{sgn}\left( \alpha_{\mathrm{Y}}\right)  \sqrt{-\kappa_{4}\omega_{\mathrm{eff}}\rho_{\mathrm{eff}}},
\end{equation}
and using the estimated current dark matter density in the Milky Way halo as $\rho_{\mathrm{eff}}$ (around $0.4\, \mathrm{GeV}/\mathrm{cm}^{3}$, see refs.~\cite{galaxies8020037,Pillepich_2014}), we find
\begin{equation}
\frac{\mathrm{d}\Theta_{\mathrm{grav}}}{\mathrm{d}\ell}\sim \pm\frac{1^\circ}{5\times10^{6}\,\mathrm{ly}},
\end{equation}
indicating that the effect is indeed very small. As a reference, the Local Group has a diameter of $\times10^{7}\,\mathrm{ly}$. When we use the current average dark matter energy density in the universe as $\rho_{\mathrm{eff}}$, approximately $4\, \mathrm{keV}/\mathrm{cm}^{3}$, the angle decreases by three orders of magnitude.

However, the rate of this gravitational Faraday effect increases significantly in the early universe. In terms of redshift, the effective density evolves as
\begin{equation}
\rho_{\mathrm{eff}}\left(  z\right)  =\rho_{\mathrm{eff}0}\left(  1+z\right)^{3\left(  1+\omega_{\mathrm{eff}}\right)},
\end{equation}
and the rate of change in the rotation angle becomes
\begin{equation}
\left.  \frac{\mathrm{d}\Theta_{\mathrm{grav}}}{\mathrm{d}\ell}\right|_{z}=\left.  \frac{\mathrm{d}\Theta_{\mathrm{grav}}}{\mathrm{d}\ell}\right| _{0}\left(  1+z\right)^{\frac{3}{2}\left(  1+\omega_{\mathrm{eff}}\right)}.
\end{equation}

Therefore, at the CMB epoch ($z=1089$) the rotation rate is approximately $3.2\times10^{4}$ times faster than at $z=0$. To higher $z$ (e.g., primordial GWs) the effect would be even bigger.

For other ECSK models, such as the one in ref.~\cite{Yun:2024xzk} that links torsion to dark energy effects, the exact values will vary. However, we generally expect:
\begin{equation}
\kappa_{4}\left[  S^{0}\right]^{2} \sim \rho_{\mathrm{D}},
\end{equation}
with the overall behavior following a similar pattern. So, if dark matter has some non-vanishing spin tensor, we could expect the gravitational Faraday effect to be much more important in the early universe.

\section*{Acknowledgments}
FI acknowledges financial support from the Chilean government through Fondecyt grant 1211219. FI is thankful of the warm hospitality at ICEN, Iquique, where most of this work was made, and of the emotional support of the Netherlands Bach Society. They made freely available superb quality recordings of the music of Bach, and without them, this work would have been impossible. The work of C.Q. is founded by Fondecyt grant 11231238 from government of Chile. OV acknowledges financial support from the Chilean government through Fondecyt 11200742.
\appendix
\section{Polarization Modes Basis}\label{App_Polarization}

When analyzing gravitational waves (GWs) within an arbitrary gravity theory that includes non-vanishing torsion in $d=4$, the only viable gauge fixing is typically the Lorenz gauge~\cite{Barrientos2019}:
\begin{equation}
\mathcal{D}_{a}H^{a}-\frac{1}{2}\mathrm{dI}_{a}H^{a}=0.
\end{equation}
In this framework, we must account for all six polarization modes of the GW, which are usually labeled in the literature as $\left\{ +,\times,b,l,x,y\right\}$. Specifically, in the case of the ECSK theory with an axial torsion background, further gauge fixing is possible, leaving only the $\left \{ +,\times \right \}$ modes, as in standard GR. However, for this work, we choose to retain all six polarization modes to demonstrate the general approach and allow the dynamics to suppress the extra modes naturally.

The choice of a polarization basis is largely arbitrary. For simplicity, we work in an orthonormal frame and rotate it such that the third spatial direction aligns with the GW propagation. With this, the basis is chosen~\cite{Nishizawa:2009jh} as:
\begin{equation}
\mathbb{P}_{mn}^{\left(  +\right)  }=\frac{1}{\sqrt{2}}\left(
\begin{array}
[c]{cccc}
0 & 0 & 0 & 0\\
0 & 1 & 0 & 0\\
0 & 0 & -1 & 0\\
0 & 0 & 0 & 0
\end{array}
\right)  ,\qquad \mathbb{P}_{mn}^{\left(  \times \right)  }=\frac{1}{\sqrt{2}}\left(
\begin{array}
[c]{cccc}%
0 & 0 & 0 & 0\\
0 & 0 & 1 & 0\\
0 & 1 & 0 & 0\\
0 & 0 & 0 & 0
\end{array}
\right) ,
\end{equation}
\begin{equation}
\mathbb{P}_{mn}^{\left( b\right) }=\frac{1}{\sqrt{2}}\left(
\begin{array}
[c]{cccc}
0 & 0 & 0 & 0\\
0 & 1 & 0 & 0\\
0 & 0 & 1 & 0\\
0 & 0 & 0 & 0
\end{array}
\right) ,\qquad \mathbb{P}_{mn}^{\left(  l\right)  }=\left(
\begin{array}
[c]{cccc}
0 & 0 & 0 & 0\\
0 & 0 & 0 & 0\\
0 & 0 & 0 & 0\\
0 & 0 & 0 & 1
\end{array}
\right) ,
\end{equation}
\begin{equation}
\mathbb{P}_{mn}^{\left(  x\right)  }=\frac{1}{\sqrt{2}}\left(
\begin{array}
[c]{cccc}
0 & 0 & 0 & 0\\
0 & 0 & 0 & 1\\
0 & 0 & 0 & 0\\
0 & 1 & 0 & 0
\end{array}
\right)  ,\qquad \mathbb{P}_{mn}^{\left(  y\right)  }=\frac{1}{\sqrt{2}}\left(
\begin{array}
[c]{cccc}
0 & 0 & 0 & 0\\
0 & 0 & 0 & 0\\
0 & 0 & 0 & 1\\
0 & 0 & 1 & 0
\end{array}
\right) .
\end{equation}

This basis satisfies the orthonormality condition in eq.~\ref{Eq_P_orthonormal}. The non-zero commutators of the polarization matrices are:
\begin{equation}
\begin{tabular}
[c]{lrl}%
$\left[  \mathbb{P}^{\left(  +\right)  },\mathbb{P}^{\left(  \times \right)}\right]  ^{ab}=\delta_{12}^{ab},$ & $\qquad$ & $\left[  \mathbb{P}^{\left(b\right)  },\mathbb{P}^{\left(  x\right)  }\right]  ^{ab}=\frac{1}{2}\delta_{13}^{ab},$\\

$\left[  \mathbb{P}^{\left(  +\right)  },\mathbb{P}^{\left(  x\right)}\right]  ^{ab}=\frac{1}{2}\delta_{13}^{ab},$ &  & $\left[  \mathbb{P}^{\left(  b\right)  },\mathbb{P}^{\left(  y\right)  }\right]  ^{ab}=\frac{1}{2}\delta_{23}^{ab},$\\

$\left[  \mathbb{P}^{\left(  +\right)  },\mathbb{P}^{\left(  y\right)}\right]  ^{ab}=-\frac{1}{2}\delta_{23}^{ab},$ &  & $\left[  \mathbb{P}^{\left(  l\right)  },\mathbb{P}^{\left(  x\right)  }\right] ^{ab}=-\frac{1}{\sqrt{2}}\delta_{13}^{ab},$\\

$\left[  \mathbb{P}^{\left(  \times \right)  },\mathbb{P}^{\left(  x\right)}\right]  ^{ab}=\frac{1}{2}\delta_{23}^{ab},$ &  & $\left[  \mathbb{P}^{\left(  l\right)  },\mathbb{P}^{\left(  y\right)  }\right]  ^{ab}=-\frac{1}{\sqrt{2}}\delta_{23}^{ab},$\\

$\left[  \mathbb{P}^{\left(  \times \right)  },\mathbb{P}^{\left(  y\right)}\right]  ^{ab}=\frac{1}{2}\delta_{13}^{ab},$ &  & $\left[  \mathbb{P}^{\left(  x\right)  },\mathbb{P}^{\left(  y\right)  }\right]  ^{ab}=\frac{1}{2}\delta_{12}^{ab},$
\end{tabular}
\end{equation}
where $\delta_{cd}^{ab}=\delta_{c}^{a}\delta_{d}^{b}-\delta_{d}^{a}\delta_{c}^{b}$.

% Bibliography

%% [A] Recommended: using JHEP.bst file
\bibliographystyle{JHEP}
%\bibliography{biblio.bib}
\bibliography{Bib_2024_June_C}

\providecommand{\href}[2]{#2}\begingroup\raggedright\begin{thebibliography}{100}

\bibitem{XENON:2023cxc}
{\scshape XENON} collaboration, \emph{{First Dark Matter Search with Nuclear
  Recoils from the XENONnT Experiment}},
  \href{https://doi.org/10.1103/PhysRevLett.131.041003}{\emph{Phys. Rev. Lett.}
  {\bfseries 131} (2023) 041003}
  [\href{https://arxiv.org/abs/2303.14729}{{\ttfamily 2303.14729}}].

\bibitem{XENON:2024ijk}
{\scshape XENON} collaboration, \emph{{First Measurement of Solar $^8$B
  Neutrinos via Coherent Elastic Neutrino-Nucleus Scattering with XENONnT}},
  \href{https://arxiv.org/abs/2408.02877}{{\ttfamily 2408.02877}}.

\bibitem{LZ:2022lsv}
{\scshape LZ} collaboration, \emph{{First Dark Matter Search Results from the
  LUX-ZEPLIN (LZ) Experiment}},
  \href{https://doi.org/10.1103/PhysRevLett.131.041002}{\emph{Phys. Rev. Lett.}
  {\bfseries 131} (2023) 041002}
  [\href{https://arxiv.org/abs/2207.03764}{{\ttfamily 2207.03764}}].

\bibitem{LZ:2024psa}
{\scshape LZ} collaboration, \emph{{New constraints on ultraheavy dark matter
  from the LZ experiment}},
  \href{https://doi.org/10.1103/PhysRevD.109.112010}{\emph{Phys. Rev. D}
  {\bfseries 109} (2024) 112010}
  [\href{https://arxiv.org/abs/2402.08865}{{\ttfamily 2402.08865}}].

\bibitem{Pandey:2024dcd}
S.~Pandey, E.D.~Hall and M.~Evans, \emph{{First Results from the Axion
  Dark-Matter Birefringent Cavity (ADBC) Experiment}},
  \href{https://doi.org/10.1103/PhysRevLett.133.111003}{\emph{Phys. Rev. Lett.}
  {\bfseries 133} (2024) 111003}
  [\href{https://arxiv.org/abs/2404.12517}{{\ttfamily 2404.12517}}].

\bibitem{Cho:2007cb}
A.~Cho, \emph{{Large hadron collider: Physicists' Nightmare Scenario: The Higgs
  and Nothing Else}},
  \href{https://doi.org/10.1126/science.315.5819.1657}{\emph{Science}
  {\bfseries 315} (2007) 1657}.

\bibitem{Bertone:2011kb}
G.~Bertone, D.~Cumberbatch, R.~Ruiz~de Austri and R.~Trotta, \emph{{Dark Matter
  Searches: The Nightmare Scenario}},
  \href{https://doi.org/10.1088/1475-7516/2012/01/004}{\emph{JCAP} {\bfseries
  01} (2012) 004} [\href{https://arxiv.org/abs/1107.5813}{{\ttfamily
  1107.5813}}].

\bibitem{Tambalo:2022wlm}
G.~Tambalo, M.~Zumalac\'arregui, L.~Dai and M.H.-Y.~Cheung,
  \emph{{Gravitational wave lensing as a probe of halo properties and dark
  matter}}, \href{https://doi.org/10.1103/PhysRevD.108.103529}{\emph{Phys. Rev.
  D} {\bfseries 108} (2023) 103529}
  [\href{https://arxiv.org/abs/2212.11960}{{\ttfamily 2212.11960}}].

\bibitem{Ezquiaga:2020dao}
J.M.~Ezquiaga and M.~Zumalac\'arregui, \emph{{Gravitational wave lensing beyond
  general relativity: birefringence, echoes and shadows}},
  \href{https://doi.org/10.1103/PhysRevD.102.124048}{\emph{Phys. Rev. D}
  {\bfseries 102} (2020) 124048}
  [\href{https://arxiv.org/abs/2009.12187}{{\ttfamily 2009.12187}}].

\bibitem{Siddhartha:2019yjm}
M.~Siddhartha and A.~Dasgupta, \emph{{Scalar and fermion field interactions
  with a gravitational wave}},
  \href{https://doi.org/10.1088/1361-6382/ab79d6}{\emph{Class. Quant. Grav.}
  {\bfseries 37} (2020) 105001}
  [\href{https://arxiv.org/abs/1907.07531}{{\ttfamily 1907.07531}}].

\bibitem{Hehl:1976kt}
F.W.~Hehl, G.D.~Kerlick and P.~Von Der~Heyde, \emph{{On Hypermomentum in
  General Relativity. 2. The Geometry of Space-Time}}, {\emph{Z. Naturforsch.
  A} {\bfseries 31} (1976) 524}.

\bibitem{Hehl:1976kv}
F.W.~Hehl, G.D.~Kerlick and P.~Von Der~Heyde, \emph{{On Hypermomentum in
  General Relativity. 3. Coupling Hypermomentum to Geometry}}, {\emph{Z.
  Naturforsch. A} {\bfseries 31} (1976) 823}.

\bibitem{Blagojevic:2013xpa}
M.~Blagojevi{\'c} and F.W.~Hehl, eds., \emph{{Gauge Theories of Gravitation}},
  World Scientific, Singapore (2013).

\bibitem{Hehl1980}
F.W.~Hehl, \emph{{Four Lectures on Poincar{\'e} Gauge Field Theory}},  in
  \emph{{Cosmology and Gravitation: Spin, Torsion, Rotation, and
  Supergravity}}, P.G.~Bergmann and V.~De~Sabbata, eds., (New York), pp.~5--62,
  Plenum Press, 1980.

\bibitem{Hehl76}
F.W.~Hehl, P.~von~der Heyde, G.D.~Kerlick and J.M.~Nester, \emph{General
  relativity with spin and torsion: Foundations and prospects},
  \href{https://doi.org/10.1103/RevModPhys.48.393}{\emph{Rev. Mod. Phys.}
  {\bfseries 48} (1976) 393}.

\bibitem{BeltranJimenez2019}
J.~Beltr\'an~Jim\'enez, L.~Heisenberg and T.S.~Koivisto, \emph{{The Geometrical
  Trinity of Gravity}},
  \href{https://doi.org/10.3390/universe5070173}{\emph{Universe} {\bfseries 5}
  (2019) 173} [\href{https://arxiv.org/abs/1903.06830}{{\ttfamily
  1903.06830}}].

\bibitem{Addazi:2024rzo}
A.~Addazi, S.~Capozziello, A.~Marciano and G.~Meluccio, \emph{{Gravity from
  Pre-geometry}},  \href{https://arxiv.org/abs/2409.02200}{{\ttfamily
  2409.02200}}.

\bibitem{Obukhov:2017pxa}
Y.N.~Obukhov, \emph{{Gravitational waves in Poincar\'e gauge gravity theory}},
  \href{https://doi.org/10.1103/PhysRevD.95.084028}{\emph{Phys. Rev. D}
  {\bfseries 95} (2017) 084028}
  [\href{https://arxiv.org/abs/1702.05185}{{\ttfamily 1702.05185}}].

\bibitem{Jimenez-Cano:2022arz}
A.~Jim\'enez-Cano, \emph{{Review of gravitational wave solutions in quadratic
  metric-affine gravity}},  in \emph{{2022 Snowmass Summer Study}}, 3, 2022
  [\href{https://arxiv.org/abs/2203.03936}{{\ttfamily 2203.03936}}].

\bibitem{Jimenez-Cano:2020chm}
A.~Jim\'enez-Cano, \emph{{New metric-affine generalizations of gravitational
  wave geometries}},
  \href{https://doi.org/10.1140/epjc/s10052-020-8239-5}{\emph{Eur. Phys. J. C}
  {\bfseries 80} (2020) 672}
  [\href{https://arxiv.org/abs/2005.02014}{{\ttfamily 2005.02014}}].

\bibitem{Jimenez-Cano:2020lea}
A.~Jim\'enez-Cano and Y.N.~Obukhov, \emph{{Gravitational waves in metric-affine
  gravity theory}},
  \href{https://doi.org/10.1103/PhysRevD.103.024018}{\emph{Phys. Rev. D}
  {\bfseries 103} (2021) 024018}
  [\href{https://arxiv.org/abs/2010.14528}{{\ttfamily 2010.14528}}].

\bibitem{Ranjbar:2024bip}
M.~Ranjbar, S.~Akhshabi and M.~Shadmehri, \emph{{Gravitational slip parameter
  and gravitational waves in Einstein\textendash{}Cartan theory}},
  \href{https://doi.org/10.1140/epjc/s10052-024-12670-4}{\emph{Eur. Phys. J. C}
  {\bfseries 84} (2024) 316}
  [\href{https://arxiv.org/abs/2401.02129}{{\ttfamily 2401.02129}}].

\bibitem{deAndrade:2021qew}
L.C.G.~de~Andrade, \emph{{Gravitational waves LIGO data sourced by spin-1 and
  spin-0 torsion propagating modes in teleparallelism}},
  \href{https://doi.org/10.1140/epjp/s13360-020-01065-5}{\emph{Eur. Phys. J.
  Plus} {\bfseries 136} (2021) 146}.

\bibitem{Bluhm:2023kph}
R.~Bluhm and Y.~Zhi, \emph{{Spontaneous and Explicit Spacetime Symmetry
  Breaking in Einstein\textendash{}Cartan Theory with Background Fields}},
  \href{https://doi.org/10.3390/sym16010025}{\emph{Symmetry} {\bfseries 16}
  (2024) 25} [\href{https://arxiv.org/abs/2312.14071}{{\ttfamily 2312.14071}}].

\bibitem{Ezquiaga:2017ner}
J.M.~Ezquiaga, J.~Garc\'\i{}a-Bellido and M.~Zumalac\'arregui, \emph{{Field
  redefinitions in theories beyond Einstein gravity using the language of
  differential forms}},
  \href{https://doi.org/10.1103/PhysRevD.95.084039}{\emph{Phys. Rev. D}
  {\bfseries 95} (2017) 084039}
  [\href{https://arxiv.org/abs/1701.05476}{{\ttfamily 1701.05476}}].

\bibitem{Blagojevic:2017ssv}
M.~Blagojevi\'c, B.~Cvetkovi\'c and Y.N.~Obukhov, \emph{{Generalized plane
  waves in Poincar\'e gauge theory of gravity}},
  \href{https://doi.org/10.1103/PhysRevD.96.064031}{\emph{Phys. Rev. D}
  {\bfseries 96} (2017) 064031}
  [\href{https://arxiv.org/abs/1708.08766}{{\ttfamily 1708.08766}}].

\bibitem{Bahamonde:2021dqn}
S.~Bahamonde, M.~Caruana, K.F.~Dialektopoulos, V.~Gakis, M.~Hohmann,
  J.~Levi~Said et~al., \emph{{Gravitational-wave propagation and polarizations
  in the teleparallel analog of Horndeski gravity}},
  \href{https://doi.org/10.1103/PhysRevD.104.084082}{\emph{Phys. Rev. D}
  {\bfseries 104} (2021) 084082}
  [\href{https://arxiv.org/abs/2105.13243}{{\ttfamily 2105.13243}}].

\bibitem{Izaurieta:2019dix}
F.~Izaurieta, E.~Rodr{\'\i}guez and O.~Valdivia, \emph{{Linear and Second-order
  Geometry Perturbations on Spacetimes with Torsion}},
  \href{https://doi.org/10.1140/epjc/s10052-019-6852-y}{\emph{Eur. Phys. J. C}
  {\bfseries 79} (2019) 337}
  [\href{https://arxiv.org/abs/1901.06400}{{\ttfamily 1901.06400}}].

\bibitem{Barrientos2019}
J.~Barrientos, F.~Cordonier-Tello, C.~Corral, F.~Izaurieta, P.~Medina,
  E.~Rodr\'\i{}guez et~al., \emph{{Luminal Propagation of Gravitational Waves
  in Scalar-tensor Theories: The Case for Torsion}},
  \href{https://doi.org/10.1103/PhysRevD.100.124039}{\emph{Phys. Rev. D}
  {\bfseries 100} (2019) 124039}
  [\href{https://arxiv.org/abs/1910.00148}{{\ttfamily 1910.00148}}].

\bibitem{Barrientos:2019msu}
J.~Barrientos, F.~Izaurieta, E.~Rodr\'\i{}guez and O.~Valdivia, \emph{{Wave
  Operators, Torsion, and Weitzenb\"ock Identities}},
  \href{https://arxiv.org/abs/1903.04712}{{\ttfamily 1903.04712}}.

\bibitem{Elizalde2023}
E.~Elizalde, F.~Izaurieta, C.~Riveros, G.~Salgado and O.~Valdivia,
  \emph{{Gravitational waves in Einstein\textendash{}Cartan theory: On the
  effects of dark matter spin tensor}},
  \href{https://doi.org/10.1016/j.dark.2023.101197}{\emph{Phys. Dark Univ.}
  {\bfseries 40} (2023) 101197}
  [\href{https://arxiv.org/abs/2204.00090}{{\ttfamily 2204.00090}}].

\bibitem{Hehl:1971vek}
F.W.~Hehl, \emph{{How does one measure torsion of space-time?}},
  \href{https://doi.org/10.1016/0375-9601(71)90433-6}{\emph{Phys. Lett. A}
  {\bfseries 36} (1971) 225}.

\bibitem{Hehl:2013qga}
F.W.~Hehl, Y.N.~Obukhov and D.~Puetzfeld, \emph{{On Poincar\'e gauge theory of
  gravity, its equations of motion, and Gravity Probe B}},
  \href{https://doi.org/10.1016/j.physleta.2013.04.055}{\emph{Phys. Lett. A}
  {\bfseries 377} (2013) 1775}
  [\href{https://arxiv.org/abs/1304.2769}{{\ttfamily 1304.2769}}].

\bibitem{Puetzfeld:2014sja}
D.~Puetzfeld and Y.N.~Obukhov, \emph{{Prospects of detecting spacetime
  torsion}}, \href{https://doi.org/10.1142/S0218271814420048}{\emph{Int. J.
  Mod. Phys. D} {\bfseries 23} (2014) 1442004}
  [\href{https://arxiv.org/abs/1405.4137}{{\ttfamily 1405.4137}}].

\bibitem{Barker:2020gcp}
W.~Barker, A.~Lasenby, M.~Hobson and W.~Handley, \emph{{Addressing $H_0$
  tension with emergent dark radiation in unitary gravity}},
  \href{https://arxiv.org/abs/2003.02690}{{\ttfamily 2003.02690}}.

\bibitem{Alexander:2019wne}
S.~Alexander, M.~Cort\^es, A.R.~Liddle, J.a.~Magueijo, R.~Sims and L.~Smolin,
  \emph{{Cosmology of minimal varying Lambda theories}},
  \href{https://doi.org/10.1103/PhysRevD.100.083507}{\emph{Phys. Rev. D}
  {\bfseries 100} (2019) 083507}
  [\href{https://arxiv.org/abs/1905.10382}{{\ttfamily 1905.10382}}].

\bibitem{Alexander:2020umk}
S.~Alexander, L.~Jenks, P.~Jirouvsek, J.~Magueijo and T.~Z{\l}o{\'s}nik,
  \emph{{Gravity waves in parity-violating Copernican Universes}},
  \href{https://arxiv.org/abs/2001.06373}{{\ttfamily 2001.06373}}.

\bibitem{Tilquin:2011bu}
A.~Tilquin and T.~Schucker, \emph{{Torsion, an alternative to dark matter?}},
  \href{https://doi.org/10.1007/s10714-011-1222-6}{\emph{Gen. Rel. Grav.}
  {\bfseries 43} (2011) 2965}
  [\href{https://arxiv.org/abs/1104.0160}{{\ttfamily 1104.0160}}].

\bibitem{Poplawski:2011jz}
N.J.~Pop\l{}awski, \emph{{Nonsingular, big-bounce cosmology from spinor-torsion
  coupling}}, \href{https://doi.org/10.1103/PhysRevD.85.107502}{\emph{Phys.
  Rev. D} {\bfseries 85} (2012) 107502}
  [\href{https://arxiv.org/abs/1111.4595}{{\ttfamily 1111.4595}}].

\bibitem{Poplawski:2011qr}
N.J.~Pop\l{}awski, \emph{{Spacetime torsion as a possible remedy to major
  problems in gravity and cosmology}},
  \href{https://doi.org/10.1080/21672857.2013.11519725}{\emph{Astron. Rev.}
  {\bfseries 8} (2013) 108} [\href{https://arxiv.org/abs/1106.4859}{{\ttfamily
  1106.4859}}].

\bibitem{Poplawski:2012qy}
N.J.~Pop\l{}awski, \emph{{Thermal fluctuations in
  Einstein-Cartan-Sciama-Kibble-Dirac bouncing cosmology}},
  \href{https://arxiv.org/abs/1201.0316}{{\ttfamily 1201.0316}}.

\bibitem{Unger:2018oqo}
G.~Unger and N.~Pop\l{}awski, \emph{{Big bounce and closed universe from spin
  and torsion}},
  \href{https://doi.org/10.3847/1538-4357/aaf169}{\emph{Astrophys. J.}
  {\bfseries 870} (2019) 78}
  [\href{https://arxiv.org/abs/1808.08327}{{\ttfamily 1808.08327}}].

\bibitem{Desai:2016}
S.~Desai and N.J.~Pop\l{}awski, \emph{{Non-parametric reconstruction of an
  inflaton potential from
  Einstein\textendash{}Cartan\textendash{}Sciama\textendash{}Kibble gravity
  with particle production}},
  \href{https://doi.org/10.1016/j.physletb.2016.02.014}{\emph{Phys. Lett. B}
  {\bfseries 755} (2016) 183}
  [\href{https://arxiv.org/abs/1510.08834}{{\ttfamily 1510.08834}}].

\bibitem{Magueijo:2019vmk}
J.~Magueijo and T.~Z{\l}o{\'s}nik, \emph{{Parity violating Friedmann
  Universes}}, \href{https://doi.org/10.1103/PhysRevD.100.084036}{\emph{Phys.
  Rev. D} {\bfseries 100} (2019) 084036}
  [\href{https://arxiv.org/abs/1908.05184}{{\ttfamily 1908.05184}}].

\bibitem{Izaurieta:2020xpk}
F.~Izaurieta, S.~Lepe and O.~Valdivia, \emph{{The Spin Tensor of Dark Matter
  and the Hubble Parameter Tension}},
  \href{https://doi.org/10.1016/j.dark.2020.100662}{\emph{Phys. Dark Univ.}
  {\bfseries 30} (2020) 100662}
  [\href{https://arxiv.org/abs/2004.13163}{{\ttfamily 2004.13163}}].

\bibitem{Poplawski:2020pie}
N.~Pop\l{}awski, \emph{{Gravitational Collapse of a Fluid with Torsion into a
  Universe in a Black Hole}},
  \href{https://doi.org/10.1134/S1063776121030092}{\emph{J. Exp. Theor. Phys.}
  {\bfseries 132} (2021) 374}
  [\href{https://arxiv.org/abs/2008.02136}{{\ttfamily 2008.02136}}].

\bibitem{Kranas:2018jdc}
D.~Kranas, C.G.~Tsagas, J.D.~Barrow and D.~Iosifidis, \emph{{Friedmann-like
  universes with torsion}},
  \href{https://doi.org/10.1140/epjc/s10052-019-6822-4}{\emph{Eur. Phys. J. C}
  {\bfseries 79} (2019) 341}
  [\href{https://arxiv.org/abs/1809.10064}{{\ttfamily 1809.10064}}].

\bibitem{Pereira:2019yhu}
S.H.~Pereira, R.d.C.~Lima, J.F.~Jesus and R.F.L.~Holanda, \emph{{Acceleration
  in Friedmann cosmology with torsion}},
  \href{https://doi.org/10.1140/epjc/s10052-019-7462-4}{\emph{Eur. Phys. J. C}
  {\bfseries 79} (2019) 950}
  [\href{https://arxiv.org/abs/1906.07624}{{\ttfamily 1906.07624}}].

\bibitem{Pereira:2022cmu}
S.H.~Pereira, A.M.~Vicente, J.F.~Jesus and R.F.L.~Holanda, \emph{{Dark matter
  from torsion in Friedmann cosmology}},
  \href{https://arxiv.org/abs/2202.01807}{{\ttfamily 2202.01807}}.

\bibitem{Guimaraes:2020drj}
T.M.~Guimar\~aes, R.d.C.~Lima and S.H.~Pereira, \emph{{Cosmological inflation
  driven by a scalar torsion function}},
  \href{https://doi.org/10.1140/epjc/s10052-021-09076-x}{\emph{Eur. Phys. J. C}
  {\bfseries 81} (2021) 271}
  [\href{https://arxiv.org/abs/2011.13906}{{\ttfamily 2011.13906}}].

\bibitem{Ivanov:2016xjm}
A.~Ivanov and M.~Wellenzohn, \emph{{Einstein--cartan Gravity with Torsion Field
  Serving as an Origin for the Cosmological Constant or Dark Energy Density}},
  \href{https://doi.org/10.3847/0004-637X/829/1/47}{\emph{Astrophys. J.}
  {\bfseries 829} (2016) 47}
  [\href{https://arxiv.org/abs/1607.01128}{{\ttfamily 1607.01128}}].

\bibitem{Razina:2010bj}
O.~Razina, Y.~Myrzakulov, N.~Serikbayev, G.~Nugmanova and R.~Myrzakulov,
  \emph{{Einstein-Cartan gravity with scalar-fermion interactions}},
  \href{https://doi.org/10.2478/s11534-011-0102-8}{\emph{Central Eur. J. Phys.}
  {\bfseries 10} (2012) 47} [\href{https://arxiv.org/abs/1012.5690}{{\ttfamily
  1012.5690}}].

\bibitem{Palle:2014goa}
D.~Palle, \emph{{On the Einstein-Cartan cosmology vs. Planck data}},
  \href{https://doi.org/10.1134/S1063776114030157}{\emph{J. Exp. Theor. Phys.}
  {\bfseries 118} (2014) 587}
  [\href{https://arxiv.org/abs/1405.3435}{{\ttfamily 1405.3435}}].

\bibitem{Rani:2022jam}
S.~Rani, I.~Ashraf and A.~Jawad, \emph{{Cosmic analysis for some parametrized
  squared speed of sound models in nonzero torsion cosmology}},
  \href{https://doi.org/10.1142/S0218271822500948}{\emph{Int. J. Mod. Phys. D}
  {\bfseries 31} (2022) 2250094}.

\bibitem{Bakry:2021ajt}
M.A.~Bakry, A.~Eid and M.M.~Khader, \emph{{Torsion and shear effect on a Big
  Rip model in a gravitational field}},
  \href{https://doi.org/10.1007/s10509-021-04002-9}{\emph{Astrophys. Space
  Sci.} {\bfseries 366} (2021) 97}.

\bibitem{Barker:2020elg}
W.E.V.~Barker, A.N.~Lasenby, M.P.~Hobson and W.J.~Handley, \emph{{Mapping
  Poincar\'e gauge cosmology to Horndeski theory for emergent dark energy}},
  \href{https://doi.org/10.1103/PhysRevD.102.084002}{\emph{Phys. Rev. D}
  {\bfseries 102} (2020) 084002}
  [\href{https://arxiv.org/abs/2006.03581}{{\ttfamily 2006.03581}}].

\bibitem{Kasem:2020wsp}
A.~Kasem and S.~Khalil, \emph{{Quantum cosmology with vector torsion}},
  \href{https://doi.org/10.1209/0295-5075/ac39ed}{\emph{EPL} {\bfseries 139}
  (2022) 19002} [\href{https://arxiv.org/abs/2012.09888}{{\ttfamily
  2012.09888}}].

\bibitem{Cabral:2020mzw}
F.~Cabral, F.S.N.~Lobo and D.~Rubiera-Garcia, \emph{{The cosmological principle
  in theories with torsion: The case of Einstein-Cartan-Dirac-Maxwell
  gravity}}, \href{https://doi.org/10.1088/1475-7516/2020/10/057}{\emph{JCAP}
  {\bfseries 10} (2020) 057}
  [\href{https://arxiv.org/abs/2004.13693}{{\ttfamily 2004.13693}}].

\bibitem{Huang:2024ujj}
Q.~Huang, H.~Huang, B.~Xu, K.~Zhang and H.~Chen, \emph{{Phase space analysis of
  the evolution of the early universe in Einstein-Cartan theory}},
  \href{https://arxiv.org/abs/2404.10400}{{\ttfamily 2404.10400}}.

\bibitem{Vicente:2024puv}
A.M.~Vicente, J.F.~Jesus and S.H.~Pereira, \emph{{Kinematic reconstruction of
  torsion as dark energy in Friedmann cosmology}},
  \href{https://arxiv.org/abs/2402.00844}{{\ttfamily 2402.00844}}.

\bibitem{Pervaiz:2023rkk}
N.~Pervaiz, N.~Azhar, A.~Jawad and S.~Rani, \emph{{Cosmic and thermodynamic
  analysis of parameterized torsion gravity}},
  \href{https://doi.org/10.1016/j.cjph.2023.12.020}{\emph{Chin. J. Phys.}
  {\bfseries 88} (2024) 110}.

\bibitem{Usman:2023nww}
M.~Usman and A.~Jawad, \emph{{Matter growth perturbations and cosmography in
  modified torsion cosmology}},
  \href{https://doi.org/10.1140/epjc/s10052-023-12122-5}{\emph{Eur. Phys. J. C}
  {\bfseries 83} (2023) 958}.

\bibitem{Liu:2023znv}
T.~Liu, Z.~Liu, J.~Wang, S.~Gong, M.~Li and S.~Cao, \emph{{Revisiting
  Friedmann-like cosmology with torsion: newest constraints from high-redshift
  observations}},
  \href{https://doi.org/10.1088/1475-7516/2023/07/059}{\emph{JCAP} {\bfseries
  07} (2023) 059} [\href{https://arxiv.org/abs/2304.06425}{{\ttfamily
  2304.06425}}].

\bibitem{Benisty:2022lhr}
D.~Benisty, A.~van~de Venn, D.~Vasak, J.~Struckmeier and H.~Stoecker,
  \emph{{Torsional dark energy}},
  \href{https://doi.org/10.1142/S0218271822420135}{\emph{Int. J. Mod. Phys. D}
  {\bfseries 31} (2022) 2242013}.

\bibitem{Benisty:2021sul}
D.~Benisty, E.I.~Guendelman, A.~van~de Venn, D.~Vasak, J.~Struckmeier and
  H.~Stoecker, \emph{{The dark side of the torsion: dark energy from
  propagating torsion}},
  \href{https://doi.org/10.1140/epjc/s10052-022-10187-2}{\emph{Eur. Phys. J. C}
  {\bfseries 82} (2022) 264}
  [\href{https://arxiv.org/abs/2109.01052}{{\ttfamily 2109.01052}}].

\bibitem{Chervon:2023gio}
S.V.~Chervon and I.V.~Fomin, \emph{{Reconstruction of Scalar-Torsion Gravity
  Theories from the Physical Potential of a Scalar Field}},
  \href{https://doi.org/10.3390/sym15020291}{\emph{Symmetry} {\bfseries 15}
  (2023) 291}.

\bibitem{deAndrade:2024yby}
L.C.G.~de~Andrade, \emph{{Dark torsion oscillations and bounces induced by
  Barbero-Immirzi and quantum correction on Einstein-Cartan gravity with very
  light inflatons}},
  \href{https://doi.org/10.1016/j.dark.2024.101478}{\emph{Phys. Dark Univ.}
  {\bfseries 44} (2024) 101478}.

\bibitem{Gao:2024gat}
Z.~Gao and L.C.G.~de~Andrade, \emph{{Chern-Simmons electrodynamics and torsion
  dark matter axions}},  \href{https://arxiv.org/abs/2404.13517}{{\ttfamily
  2404.13517}}.

\bibitem{Capozziello:2024lsz}
S.~Capozziello and M.~Shokri, \emph{{Comparing Inflationary Models in Extended
  Metric-Affine Theories of Gravity}},
  \href{https://arxiv.org/abs/2408.17415}{{\ttfamily 2408.17415}}.

\bibitem{SupergravityVanProeyen}
D.Z.~Freedman and A.V.~Proeyen, \emph{Supergravity}, Cambridge University
  Press, 1~ed. (5, 2012).

\bibitem{Carroll:1994dq}
S.M.~Carroll and G.B.~Field, \emph{{Consequences of propagating torsion in
  connection dynamic theories of gravity}},
  \href{https://doi.org/10.1103/PhysRevD.50.3867}{\emph{Phys. Rev. D}
  {\bfseries 50} (1994) 3867}
  [\href{https://arxiv.org/abs/gr-qc/9403058}{{\ttfamily gr-qc/9403058}}].

\bibitem{Jockel:2024fps}
C.~Jockel and L.~Menger, \emph{{The Effect of Torsion on Neutron Star Structure
  in Einstein-Cartan Gravity}},
  \href{https://arxiv.org/abs/2406.05851}{{\ttfamily 2406.05851}}.

\bibitem{Barrientos2017}
J.~Barrientos, F.~Cordonier-Tello, F.~Izaurieta, P.~Medina, D.~Narbona,
  E.~Rodr{\'\i}guez et~al., \emph{{Nonminimal couplings, gravitational waves,
  and torsion in Horndeski's theory}},
  \href{https://doi.org/10.1103/PhysRevD.96.084023}{\emph{Phys. Rev.}
  {\bfseries D96} (2017) 084023}
  [\href{https://arxiv.org/abs/1703.09686}{{\ttfamily 1703.09686}}].

\bibitem{Barrientos2022}
J.~Barrientos, F.~Izaurieta, E.~Rodr\'\i{}guez and O.~Valdivia, \emph{{Wave
  operators, torsion, and Weitzenb\"ock identities}},
  \href{https://doi.org/10.1007/s10714-022-02914-7}{\emph{Gen. Rel. Grav.}
  {\bfseries 54} (2022) 26} [\href{https://arxiv.org/abs/1903.04712}{{\ttfamily
  1903.04712}}].

\bibitem{Maggiore:1900zz}
M.~Maggiore, \emph{{Gravitational Waves. Vol. 1: Theory and Experiments}},
  Oxford Master Series in Physics, Oxford University Press (2007).

\bibitem{IceCubeColb2024}
T.I.~Collaboration, \emph{{Search for decoherence from quantum gravity with
  atmospheric neutrinos}},
  \href{https://doi.org/10.1038/s41567-024-02436-w}{\emph{Nature Phys.}
  {\bfseries 20} (2024) 913}.

\bibitem{Lee2018}
J.-W.~Lee, \emph{{Brief History of Ultra-light Scalar Dark Matter Models}},
  \href{https://doi.org/10.1051/epjconf/201816806005}{\emph{EPJ Web Conf.}
  {\bfseries 168} (2018) 06005}
  [\href{https://arxiv.org/abs/1704.05057}{{\ttfamily 1704.05057}}].

\bibitem{Hui2017}
L.~Hui, J.P.~Ostriker, S.~Tremaine and E.~Witten, \emph{{Ultralight scalars as
  cosmological dark matter}},
  \href{https://doi.org/10.1103/PhysRevD.95.043541}{\emph{Phys. Rev. D}
  {\bfseries 95} (2017) 043541}
  [\href{https://arxiv.org/abs/1610.08297}{{\ttfamily 1610.08297}}].

\bibitem{Berezhiani2015}
L.~Berezhiani and J.~Khoury, \emph{{Theory of dark matter superfluidity}},
  \href{https://doi.org/10.1103/PhysRevD.92.103510}{\emph{Phys. Rev. D}
  {\bfseries 92} (2015) 103510}
  [\href{https://arxiv.org/abs/1507.01019}{{\ttfamily 1507.01019}}].

\bibitem{Khoury2016}
J.~Khoury, \emph{{Another Path for the Emergence of Modified Galactic Dynamics
  from Dark Matter Superfluidity}},
  \href{https://doi.org/10.1103/PhysRevD.93.103533}{\emph{Phys. Rev. D}
  {\bfseries 93} (2016) 103533}
  [\href{https://arxiv.org/abs/1602.05961}{{\ttfamily 1602.05961}}].

\bibitem{Hossenfelder2017}
S.~Hossenfelder, \emph{{Covariant version of Verlinde\textquoteright{}s
  emergent gravity}},
  \href{https://doi.org/10.1103/PhysRevD.95.124018}{\emph{Phys. Rev. D}
  {\bfseries 95} (2017) 124018}
  [\href{https://arxiv.org/abs/1703.01415}{{\ttfamily 1703.01415}}].

\bibitem{Hossenfelder2019}
S.~Hossenfelder and T.~Mistele, \emph{{Strong lensing with superfluid dark
  matter}}, \href{https://doi.org/10.1088/1475-7516/2019/02/001}{\emph{JCAP}
  {\bfseries 02} (2019) 001}
  [\href{https://arxiv.org/abs/1809.00840}{{\ttfamily 1809.00840}}].

\bibitem{Mistele2022}
T.~Mistele, S.~McGaugh and S.~Hossenfelder, \emph{{Galactic mass-to-light
  ratios with superfluid dark matter}},
  \href{https://doi.org/10.1051/0004-6361/202243216}{\emph{Astron. Astrophys.}
  {\bfseries 664} (2022) A40}
  [\href{https://arxiv.org/abs/2201.07282}{{\ttfamily 2201.07282}}].

\bibitem{Vagnozzi:2023nrq}
S.~Vagnozzi, \emph{{Seven Hints That Early-Time New Physics Alone Is Not
  Sufficient to Solve the Hubble Tension}},
  \href{https://doi.org/10.3390/universe9090393}{\emph{Universe} {\bfseries 9}
  (2023) 393} [\href{https://arxiv.org/abs/2308.16628}{{\ttfamily
  2308.16628}}].

\bibitem{Vagnozzi:2019ezj}
S.~Vagnozzi, \emph{{New physics in light of the $H_0$ tension: An alternative
  view}}, \href{https://doi.org/10.1103/PhysRevD.102.023518}{\emph{Phys. Rev.
  D} {\bfseries 102} (2020) 023518}
  [\href{https://arxiv.org/abs/1907.07569}{{\ttfamily 1907.07569}}].

\bibitem{Koivisto:2023epd}
T.~Koivisto, \emph{{Cosmology in the Lorentz gauge theory}},
  \href{https://doi.org/10.1142/S0219887824500403}{\emph{Int. J. Geom. Meth.
  Mod. Phys.} {\bfseries 20} (2023) 2450040}
  [\href{https://arxiv.org/abs/2306.00963}{{\ttfamily 2306.00963}}].

\bibitem{Poplawski:2010kb}
N.J.~Pop\l{}awski, \emph{{Cosmology with torsion: An alternative to cosmic
  inflation}},
  \href{https://doi.org/10.1016/j.physletb.2010.09.056}{\emph{Phys. Lett. B}
  {\bfseries 694} (2010) 181}
  [\href{https://arxiv.org/abs/1007.0587}{{\ttfamily 1007.0587}}].

\bibitem{Boos:2016cey}
J.~Boos and F.W.~Hehl, \emph{{Gravity-induced four-fermion contact interaction
  implies gravitational intermediate W and Z type gauge bosons}},
  \href{https://doi.org/10.1007/s10773-016-3216-3}{\emph{Int. J. Theor. Phys.}
  {\bfseries 56} (2017) 751}
  [\href{https://arxiv.org/abs/1606.09273}{{\ttfamily 1606.09273}}].

\bibitem{PhysRevLett.119.161101}
{\scshape LIGO Scientific and Virgo} collaboration, \emph{{GW170817:
  Observation of Gravitational Waves from a Binary Neutron Star Inspiral}},
  \href{https://doi.org/10.1103/PhysRevLett.119.161101}{\emph{Phys. Rev. Lett.}
  {\bfseries 119} (2017) 161101}.

\bibitem{GBM:2017lvd}
{\scshape LIGO Scientific, Virgo, Fermi GBM, INTEGRAL, IceCube, AstroSat
  Cadmium Zinc Telluride Imager Team, IPN, Insight-Hxmt, ANTARES, Swift, AGILE
  Team, 1M2H Team, Dark Energy Camera GW-EM, DES, DLT40, GRAWITA, Fermi-LAT,
  ATCA, ASKAP, Las Cumbres Observatory Group, OzGrav, DWF (Deeper Wider Faster
  Program), AST3, CAASTRO, VINROUGE, MASTER, J-GEM, GROWTH, JAGWAR,
  CaltechNRAO, TTU-NRAO, NuSTAR, Pan-STARRS, MAXI Team, TZAC Consortium, KU,
  Nordic Optical Telescope, ePESSTO, GROND, Texas Tech University, SALT Group,
  TOROS, BOOTES, MWA, CALET, IKI-GW Follow-up, H.E.S.S., LOFAR, LWA, HAWC,
  Pierre Auger, ALMA, Euro VLBI Team, Pi of Sky, Chandra Team at McGill
  University, DFN, ATLAS Telescopes, High Time Resolution Universe Survey,
  RIMAS, RATIR, SKA South Africa/MeerKAT} collaboration, \emph{{Multi-messenger
  Observations of a Binary Neutron Star Merger}},
  \href{https://doi.org/10.3847/2041-8213/aa91c9}{\emph{Astrophys. J.}
  {\bfseries 848} (2017) L12}
  [\href{https://arxiv.org/abs/1710.05833}{{\ttfamily 1710.05833}}].

\bibitem{Monitor:2017mdv}
{\scshape LIGO Scientific, Virgo, Fermi-GBM, INTEGRAL} collaboration,
  \emph{{Gravitational Waves and Gamma-rays from a Binary Neutron Star Merger:
  GW170817 and GRB 170817A}},
  \href{https://doi.org/10.3847/2041-8213/aa920c}{\emph{Astrophys. J.}
  {\bfseries 848} (2017) L13}
  [\href{https://arxiv.org/abs/1710.05834}{{\ttfamily 1710.05834}}].

\bibitem{Huerta2019}
E.A.~Huerta, G.~Allen, I.~Andreoni, J.M.~Antelis, E.~Bachelet, G.B.~Berriman
  et~al., \emph{Enabling real-time multi-messenger astrophysics discoveries
  with deep learning},
  \href{https://doi.org/10.1038/s42254-019-0097-4}{\emph{Nature Reviews
  Physics} {\bfseries 1} (2019) 600}.

\bibitem{Bettoni:2016mij}
D.~Bettoni, J.M.~Ezquiaga, K.~Hinterbichler and M.~Zumalac{\'a}rregui,
  \emph{{Speed of Gravitational Waves and the Fate of Scalar-Tensor Gravity}},
  \href{https://doi.org/10.1103/PhysRevD.95.084029}{\emph{Phys. Rev.}
  {\bfseries D95} (2017) 084029}
  [\href{https://arxiv.org/abs/1608.01982}{{\ttfamily 1608.01982}}].

\bibitem{Ezquiaga:2018btd}
J.M.~Ezquiaga and M.~Zumalac{\'a}rregui, \emph{{Dark Energy in light of
  Multi-Messenger Gravitational-Wave astronomy}},
  \href{https://doi.org/10.3389/fspas.2018.00044}{\emph{Front. Astron. Space
  Sci.} {\bfseries 5} (2018) 44}
  [\href{https://arxiv.org/abs/1807.09241}{{\ttfamily 1807.09241}}].

\bibitem{Battista:2023znv}
E.~Battista, V.~De~Falco and D.~Usseglio, \emph{{First post-Newtonian N-body
  problem in Einstein\textendash{}Cartan theory with the Weyssenhoff fluid:
  Lagrangian and first integrals}},
  \href{https://doi.org/10.1140/epjc/s10052-023-11249-9}{\emph{Eur. Phys. J. C}
  {\bfseries 83} (2023) 112}
  [\href{https://arxiv.org/abs/2301.08954}{{\ttfamily 2301.08954}}].

\bibitem{Battista:2022sci}
E.~Battista and V.~De~Falco, \emph{{First post-Newtonian N-body problem in
  Einstein\textendash{}Cartan theory with the Weyssenhoff fluid: equations of
  motion}}, \href{https://doi.org/10.1140/epjc/s10052-022-10746-7}{\emph{Eur.
  Phys. J. C} {\bfseries 82} (2022) 782}
  [\href{https://arxiv.org/abs/2208.09839}{{\ttfamily 2208.09839}}].

\bibitem{Battista:2022hmv}
E.~Battista and V.~De~Falco, \emph{{Gravitational waves at the first
  post-Newtonian order with the Weyssenhoff fluid in
  Einstein\textendash{}Cartan theory}},
  \href{https://doi.org/10.1140/epjc/s10052-022-10558-9}{\emph{Eur. Phys. J. C}
  {\bfseries 82} (2022) 628}
  [\href{https://arxiv.org/abs/2206.12907}{{\ttfamily 2206.12907}}].

\bibitem{Battista:2021rlh}
E.~Battista and V.~De~Falco, \emph{{First post-Newtonian generation of
  gravitational waves in Einstein-Cartan theory}},
  \href{https://doi.org/10.1103/PhysRevD.104.084067}{\emph{Phys. Rev. D}
  {\bfseries 104} (2021) 084067}
  [\href{https://arxiv.org/abs/2109.01384}{{\ttfamily 2109.01384}}].

\bibitem{FaradaysDiary4}
M.~Faraday, \emph{Faraday's Diary}, vol.~4, HR Direct (2009).

\bibitem{Prati_Faraday:2003}
E.~Prati, \emph{Propagation in gyroelectromagnetic guiding systems},
  \href{https://doi.org/10.1163/156939303322519810}{\emph{Journal of
  Electromagnetic Waves and Applications} {\bfseries 17} (2003) 1177}.

\bibitem{PhysRevD.100.064028}
S.~Hou, X.-L.~Fan and Z.-H.~Zhu, \emph{Gravitational lensing of gravitational
  waves: Rotation of polarization plane},
  \href{https://doi.org/10.1103/PhysRevD.100.064028}{\emph{Phys. Rev. D}
  {\bfseries 100} (2019) 064028}.

\bibitem{Harikumar:2023gzh}
S.~Harikumar, L.~J\"arv, M.~Saal, A.~Wojnar and M.~Biesiada, \emph{{Propagation
  and lensing of gravitational waves in Palatini $f(\hat{R})$ gravity}},
  \href{https://doi.org/10.1103/PhysRevD.109.124014}{\emph{Phys. Rev. D}
  {\bfseries 109} (2024) 124014}
  [\href{https://arxiv.org/abs/2312.09908}{{\ttfamily 2312.09908}}].

\bibitem{Morawetz:2023jll}
K.~Morawetz, \emph{{Time behavior of Hubble parameter by torsion}},
  \href{https://doi.org/10.1142/S0217732323501924}{\emph{Mod. Phys. Lett. A}
  {\bfseries 39} (2024) 2350192}
  [\href{https://arxiv.org/abs/2303.10356}{{\ttfamily 2303.10356}}].

\bibitem{Li:2023ubd}
S.~Li and Y.~Chen, \emph{{Reconstructing Torsion Cosmology from Interacting
  Holographic Dark Energy Model}},
  \href{https://doi.org/10.3390/universe9020100}{\emph{Universe} {\bfseries 9}
  (2023) 100}.

\bibitem{Yun:2024xzk}
Y.~Yun and J.~Lee, \emph{{Holographic Dark Energy with Torsion}},
  \href{https://arxiv.org/abs/2407.16992}{{\ttfamily 2407.16992}}.

\bibitem{Izaurieta:2020kuy}
F.~Izaurieta, P.~Medina, N.~Merino, P.~Salgado and O.~Valdivia, \emph{{Mimetic
  Einstein-Cartan-Sciama-Kibble (ECSK) gravity}},
  \href{https://doi.org/10.1007/JHEP10(2020)150}{\emph{JHEP} {\bfseries 10}
  (2020) 150} [\href{https://arxiv.org/abs/2007.07226}{{\ttfamily
  2007.07226}}].

\bibitem{Cid:2017wtf}
A.~Cid, F.~Izaurieta, G.~Leon, P.~Medina and D.~Narbona, \emph{{Non-minimally
  coupled scalar field cosmology with torsion}},
  \href{https://doi.org/10.1088/1475-7516/2018/04/041}{\emph{JCAP} {\bfseries
  1804} (2018) 041} [\href{https://arxiv.org/abs/1704.04563}{{\ttfamily
  1704.04563}}].

\bibitem{Cruz:2020hkh}
M.~Cruz, F.~Izaurieta and S.~Lepe, \emph{{Non-zero torsion and late
  cosmology}}, \href{https://doi.org/10.1140/epjc/s10052-020-8128-y}{\emph{Eur.
  Phys. J. C} {\bfseries 80} (2020) 559}
  [\href{https://arxiv.org/abs/2005.04550}{{\ttfamily 2005.04550}}].

\bibitem{Izaurieta:2020vyh}
F.~Izaurieta and S.~Lepe, \emph{{Cosmological Dark Matter Amplification through
  Dark Torsion}}, \href{https://doi.org/10.1088/1361-6382/abb2d2}{\emph{Class.
  Quant. Grav.} {\bfseries 37} (2020) 205004}
  [\href{https://arxiv.org/abs/2004.06058}{{\ttfamily 2004.06058}}].

\bibitem{Poplawski:2019tub}
N.J.~Pop\l{}awski, \emph{{Non-particle dark matter from Hubble parameter}},
  \href{https://doi.org/10.1140/epjc/s10052-019-7230-5}{\emph{Eur. Phys. J. C}
  {\bfseries 79} (2019) 734}
  [\href{https://arxiv.org/abs/1906.03947}{{\ttfamily 1906.03947}}].

\bibitem{galaxies8020037}
Y.~Sofue, \emph{Rotation curve of the milky way and the dark matter density},
  \href{https://doi.org/10.3390/galaxies8020037}{\emph{Galaxies} {\bfseries 8}
  (2020) }.

\bibitem{Pillepich_2014}
A.~Pillepich, M.~Kuhlen, J.~Guedes and P.~Madau, \emph{The distribution of dark
  matter in the milky way’s disk},
  \href{https://doi.org/10.1088/0004-637X/784/2/161}{\emph{The Astrophysical
  Journal} {\bfseries 784} (2014) 161}.

\bibitem{Nishizawa:2009jh}
A.~Nishizawa, A.~Taruya and S.~Kawamura, \emph{{Cosmological test of gravity
  with polarizations of stochastic gravitational waves around 0.1-1 Hz}},
  \href{https://doi.org/10.1103/PhysRevD.81.104043}{\emph{Phys. Rev. D}
  {\bfseries 81} (2010) 104043}
  [\href{https://arxiv.org/abs/0911.0525}{{\ttfamily 0911.0525}}].

\end{thebibliography}\endgroup
%% or
%% [B] Manual formatting (see below)
%% (i) We suggest to always provide author, title and journal data or doi:
%% in short all the informations that clearly identify a document.
%% (ii) please avoid comments such as "For a review'', "For some examples",
%% "and references therein" or move them in the text. In general, please leave only references in the bibliography and move all
%% accessory text in footnotes.
%% (iii) Also, please have only one work for each \bibitem.

\end{document}